\pdfoutput=1
\documentclass[11pt,twoside,a4paper,cmspaper,final,collab]{cms-tdr}
\def\svnVersion{0a21dff}\def\svnDate{2021/11/16}\def\cmsCernNoTag{CERN-EP-2021-044}\def\cmsCernDate{\today}\def\cmsMessage{Published in the Journal of High Energy Physics as \href{http://dx.doi.org/10.1007/JHEP12(2021)106}{\doi{10.1007/JHEP12(2021)106}.}}
\begin{document}\cmsNoteHeader{B2G-19-003}

\newcommand{\bst}{\ensuremath{\PQb^\ast}\xspace}
\newcommand{\bpr}{\ensuremath{\PB}\xspace}
\newcommand{\tpr}{\ensuremath{\PQtpr}\xspace}
\newcommand{\bstr}{\ensuremath{\PQb^\ast_{\mathrm{RH}}}\xspace}
\newcommand{\bstl}{\ensuremath{\PQb^\ast_{\mathrm{LH}}}\xspace}
\newcommand{\bstlr}{\ensuremath{\PQb^\ast_{\mathrm{LH+RH}}}\xspace}
\newcommand{\tw}{\ensuremath{\PQt\PW}\xspace}
\newcommand{\mtw}{\ensuremath{m_{\tw}}\xspace}
\newcommand{\mjj}{\ensuremath{m_{\mathrm{jj}}}\xspace}
\newcommand{\mtt}{\ensuremath{m_{\PQt\PQt}}\xspace}
\newcommand{\mt}{\ensuremath{m_{\PQt}}\xspace}
\newcommand{\mW}{\ensuremath{m_{\PW}}\xspace}
\newcommand{\rpfmc}{\ensuremath{R^\mathrm{MC}_\mathrm{P/F}(\mt,\mtw)}\xspace}
\newcommand{\rpfdata}{\ensuremath{R^\text{data}_\mathrm{P/F}(\mt,\mtw)}\xspace}
\newcommand{\rrat}{\ensuremath{R_\text{ratio}(\mt,\mtw)}\xspace}
\newcommand{\rratsr}{\ensuremath{R_\text{ratio}^\mathrm{SR}(\mt,\mtw)}\xspace}
\newcommand{\rrattt}{\ensuremath{R_\text{ratio}^{\ttbar\mathrm{MR}}(\mt,\mtt)}\xspace}
\newcommand{\rpfmctt}{\ensuremath{R^\mathrm{MC}_\mathrm{P/F}(\mt,\mtt)}\xspace}
\newcommand{\rpfdatatt}{\ensuremath{R^\text{data}_\mathrm{P/F}(\mt,\mtt)}\xspace}
\providecommand{\cmsTable}[1]{\resizebox{\textwidth}{!}{#1}}

\newcolumntype{C}[1]{>{\centering\arraybackslash}p{#1}}

\cmsNoteHeader{B2G-19-003}
\title{Search for a heavy resonance decaying to a top quark and a \PW boson at \texorpdfstring{$\sqrt{s} =  13\TeV$}{sqrt(s) = 13 TeV} in the fully hadronic final state}

\date{\today}

\abstract{
A search for a heavy resonance decaying to a top quark and a \PW boson in the fully hadronic final state is presented. The analysis is performed using data from proton-proton collisions at a center-of-mass energy of 13\TeV, corresponding to an integrated luminosity of 137\fbinv recorded by the CMS experiment at the LHC. The search is focused on heavy resonances, where the decay products of each top quark or \PW boson are expected to be reconstructed as a single, large-radius jet with a distinct substructure. The production of an excited bottom quark, \bst, is used as a benchmark when setting limits on the cross section for a heavy resonance decaying to a top quark and a \PW boson. The hypotheses of \bst quarks with left-handed, right-handed, and vector-like chiralities are excluded at 95\% confidence level for masses below 2.6, 2.8, and 3.1\TeV, respectively. These are the most stringent limits on the \bst quark mass to date, extending the previous best limits by almost a factor of two.
}

\hypersetup{%
pdfauthor={CMS Collaboration},%
pdftitle={Search for a heavy resonance decaying to a top quark and a W boson at sqrt(s) = 13 TeV in the fully hadronic final state},%
pdfsubject={CMS},%
pdfkeywords={CMS, single top resonances, substructure}}

\maketitle 

\section{Introduction}
\label{sec:introduction}

The standard model (SM) has been extensively verified by experiment, nonetheless
there exists evidence that the SM is only an effective theory. Many possibilities
for physics beyond the SM have been proposed,
including the possibility that quarks are composite. Such
quarks would have an internal structure that, excited, could
produce a state with higher mass~\cite{PhysRevD.42.815,Tait:2000sh}.
Such a phenomenon is predicted by Randall--Sundrum models~\cite{RS1,RS2} and
models with a heavy gluon partner~\cite{HGP1,HGP2,HGP3}.

In this paper, we search for a heavy resonance decaying to a top quark \PQt
and a \PW boson in the fully hadronic final state, using
proton-proton ($\Pp\Pp$) collision data at a center-of-mass energy of 13\TeV.
The search uses data corresponding to an integrated luminosity of 137\fbinv
recorded by the CMS experiment~\cite{Collaboration_2008} at the CERN LHC during 2016--2018.

As a benchmark resonance, we consider an excited bottom quark, referred to as a \bst quark~\cite{Tait:2000sh}.
The strong interaction is the dominant production mechanism and can produce a single \bst quark at the LHC via
the collision of a bottom quark and a gluon, $\PQb\Pg \to \bst$. The interaction is described by the Lagrangian
\begin{linenomath}
    \begin{equation}
            \mathcal{L}_{1} = \frac{g_\text{s}}{2\Lambda}G_{\mu\nu} \overline{\PQb} \sigma^{\mu\nu} (\kappa^{\PQb}_{\text{L}}P_{\text{L}} + \kappa^{\PQb}_{\text{R}}P_{\text{R}})\bst + \text{h.c.},
            \label{eqn:Lag1}
    \end{equation}
\end{linenomath}
where $g_\text{s}$ is the strong coupling, $G_{\mu\nu}$ is the gauge field tensor
of the gluon, $\overline{\PQb}$ is the bottom quark field, $\sigma^{\mu\nu}$ is the Pauli spin
matrix, \bst is the excited bottom quark
field, and $\Lambda$ is the scale of compositeness~\cite{PhysRevD.42.815},
which is chosen to be the mass of the \bst quark. The chiral projection operators
are represented as $P_\text{L}$ and $P_\text{R}$, and $\kappa^{\PQb}_{\text{L}}$
and $\kappa^{\PQb}_{\text{R}}$ are the relative coupling strengths~\cite{Nutter_2012}.

The $\bst \to \tw$ decay is the dominant decay channel, with a branching fraction
of approximately 40\% for a \bst quark with $m_{\bst} > 1.2 \TeV$~\cite{Nutter_2012}.
The decay takes place
through the weak interaction and is described by the Lagrangian
\begin{linenomath}
    \begin{equation}
            \mathcal{L}_{2} = \frac{g_\text{2}}{\sqrt{2}}\PW^{+}_{\mu} \cPaqt \gamma^{\mu}(g_{\text{L}}P_\text{L} + g_{\text{R}}P_\text{R})\bst + \text{h.c.},
            \label{eqn:Lag2}
    \end{equation}
\end{linenomath}
where $g_\text{2}$ is the $SU(2)_L$ weak coupling and $g_\text{L}$ and $g_\text{R}$ are the
relative couplings of the \PW boson to the left- and right-handed \bst quark, respectively~\cite{Nutter_2012}.
The full interaction chain is then $\PQb\Pg \to \bst \to \PQt\PW$. The \bst quark width
is expected to be less than 10\% of the \bst quark mass, which leads to a distinct resonant structure in the mass spectrum.

Three hypotheses for the left- and right-handed \bst quark couplings are considered:
\begin{linenomath}
    \begin{alignat}{2}
        &\text{left-handed (LH): }&&\kappa^{\PQb}_{\text{L}}=g_{\text{L}}=1 \text{ and } \kappa^{\PQb}_{\text{R}}=g_{\text{R}}=0,\label{eqn:couplingsL}\\
        &\text{right-handed (RH): }&&\kappa^{\PQb}_{\text{L}}=g_{\text{L}}=0 \text{ and } \kappa^{\PQb}_{\text{R}}=g_{\text{R}}=1,\text{ and}\label{eqn:couplingsR}\\
        &\text{vector-like (LH+RH): }&&\kappa^{\PQb}_{\text{L}}=g_{\text{L}}=1 \text{ and } \kappa^{\PQb}_{\text{R}}=g_{\text{R}}=1.\label{eqn:couplings}
    \end{alignat}
\end{linenomath}
Searches for the \bst quark in the $\tw$ decay mode have been performed at the LHC by the
ATLAS Collaboration at $\sqrt{s} = 7\TeV$ ~\cite{Aad:2013rna} and by the CMS Collaboration at 8\TeV~\cite{Khachatryan2016}.
Additionally, searches for a \bst quark decaying to a bottom quark and a gluon were conducted
by the CMS Collaboration at 8\TeV~\cite{bstarTobg} and by the ATLAS Collaboration at 13\TeV~\cite{bstarTobgATLAS}. The CMS $\tw$ decay mode search
included a combination of fully hadronic, lepton+jets, and dilepton final states, and excluded
\bst quark masses at 95\% confidence level ($\CL$) below 1.4, 1.4, and 1.5\TeV, for the left-handed, right-handed, and
vector-like hypotheses, respectively.

Given the range of these exclusions, the present analysis considers a \bst quark with
a mass greater than 1.2\TeV. For these mass values, the top quark and the \PW boson
are commonly produced with a high Lorentz boost.  Because of this, the
hadronic decay products of the top quark and the \PW boson can each merge,
resulting in two massive, large-radius jets, referred to as a ``top jet'' and a ``\PW jet'', respectively.
These jets have a distinct substructure that is used to discriminate them from the
background~\cite{Larkoski_2020,Kogler_2019}.  The \bst quark mass is reconstructed as the invariant
mass of the top jet and \PW jet system, $\mtw$.  This variable, along with the
reconstructed top jet mass, $\mt$, is used to search for the \bst quark resonance.

The background is dominated by jets produced through the strong interaction, referred
to as quantum chromodynamics (QCD) multijet production, and is estimated
using multijet-enriched control regions based on inverting the top jet selection criteria. The SM {\PW}+jets and {\PZ}+jets
production backgrounds are also accounted for with this technique.
The \ttbar background is estimated with simulation templates fit to data simultaneously
in the signal region and a dedicated control region enhanced in \ttbar production that constrains the simulation templates.

A binned maximum likelihood fit to data is performed in the two-dimensional
$\mtw$ versus $\mt$ distribution, in a process where the signal and background
models are fit simultaneously. From this fit, \bst quark
mass limits are derived for the three \bst chirality hypotheses expressed
in Eqs.~(\ref{eqn:couplingsL}),~(\ref{eqn:couplingsR}), and~(\ref{eqn:couplings}).

In addition, we interpret the results under the hypothesis of a singly produced
$\bpr$ singlet vector-like quark~\cite{VLQ1,VLQ2} decaying into $\tw$. For $\bpr$ quark masses
above 1.2\TeV, the decay products would be heavily boosted with a similar signature to the \bst quark decay described above.
In the model considered, the $V_{\PQt \bpr}^\text{L}$ mixing parameter defined in Ref.~\cite{VLQ1} is set to unity,
which results in a relative signal width of less than 5\% in the B mass range of interest,
with a branching fraction to tW of approximately 50\%.
In contrast to the \bst model, the $\bpr$ quark would be produced via an electroweak interaction in association with a top or bottom quark.
We consider both scenarios, but typically the associated top or bottom quark has a much lower transverse momentum than the $\bpr$ quark decay products,
thus the effect of either on the analysis is small.

\section{The CMS detector}
\label{sec:detector}

The central feature of the CMS apparatus is a superconducting solenoid of
6\unit{m} internal diameter, providing a magnetic field of 3.8\unit{T}.
Within the solenoid volume are a silicon pixel and strip tracker, a lead
tungstate crystal electromagnetic calorimeter (ECAL), and a brass and
scintillator hadron calorimeter (HCAL), each composed of a barrel and
two endcap sections. Forward calorimeters extend the pseudorapidity
coverage provided by the barrel and endcap detectors. Muons are detected
in gas-ionization chambers embedded in the steel flux-return yoke outside the solenoid.
A more detailed description of the CMS detector, together with a definition of
the coordinate system used and the relevant kinematic variables, can be found in
Ref.~\cite{Collaboration_2008}.

Events of interest are selected using a two-tiered trigger system~\cite{Khachatryan:2016bia}.
The first level, composed of custom
hardware processors, uses information from the calorimeters and muon
detectors to select events at a rate of around 100\unit{kHz} within a fixed latency of about 4\mus. The second level, known as the
high-level trigger, consists of a farm of processors running
a version of the full event reconstruction software optimized for
fast processing, and reduces the event rate to around 1\unit{kHz} before data storage.

The analysis reflects the fact that the pixel detector was changed in the winter of 2016/2017.
The newer detector increased the number of barrel layers from three to
four and decreased the distance of the innermost layer from the beamline in
order to improve the vertex reconstruction.

\section{Data and simulated samples}
\label{sec:dataset}

CMS data taking operates on annual cycles, and thus data collection
and simulation performance can change from year to year. Therefore, we
categorize both the data and simulation by year
and apply dedicated scale factors before combining the distributions
from all three years to derive the final result.

We analyze events from the 2016 data set recorded by a trigger
that requires the scalar sum of transverse momenta, \pt, of all jets in the event,
\HT, to be at least 800 or 900\GeV, or the presence of a jet
with $\pt > 450\GeV$.
For 2017 and 2018 data, we analyze events recorded by a trigger that requires
a minimum \HT of 1050\GeV or the presence of a jet with $\pt > 500\GeV$.
Additionally, 2018 data events are recorded by a trigger that requires a jet with
$\pt > 400\GeV$ with a mass of at least 30\GeV, where the jet
trimming algorithm~\cite{Krohn_2010} has been used to reconstruct the jet mass at the
trigger level. This trigger did not exist for the 2016 or 2017 data collection,
but the addition of events recorded by this trigger provides a higher overall
selection efficiency at lower \HT for 2018.
The choice of higher \HT and jet \pt thresholds used for 2017 and 2018 are due to
an increase in the instantaneous luminosity
of the LHC between 2016 and 2017.
The combination of these triggers is nearly fully efficient for
$\mtw > 1200\GeV$.

The efficiency of the trigger selection is measured in data
as the ratio of the number of events recorded by
the combined triggers to the number of events recorded by a trigger that requires a
muon candidate with $\pt > 50\GeV$. A muon trigger is used for this measurement
because it is largely uncorrelated with the triggers used for data taking.

The trigger efficiencies are parameterized
as a function of dijet invariant mass ($\mjj$) and both the numerator and denominator of the ratio
include events that pass the preselection described in Section~\ref{sec:selection}.
The uncertainty assigned to the efficiency measurement is one half of the trigger inefficiency.

\begin{figure}[htbp]
    \centering
        \includegraphics[width=0.75\textwidth]{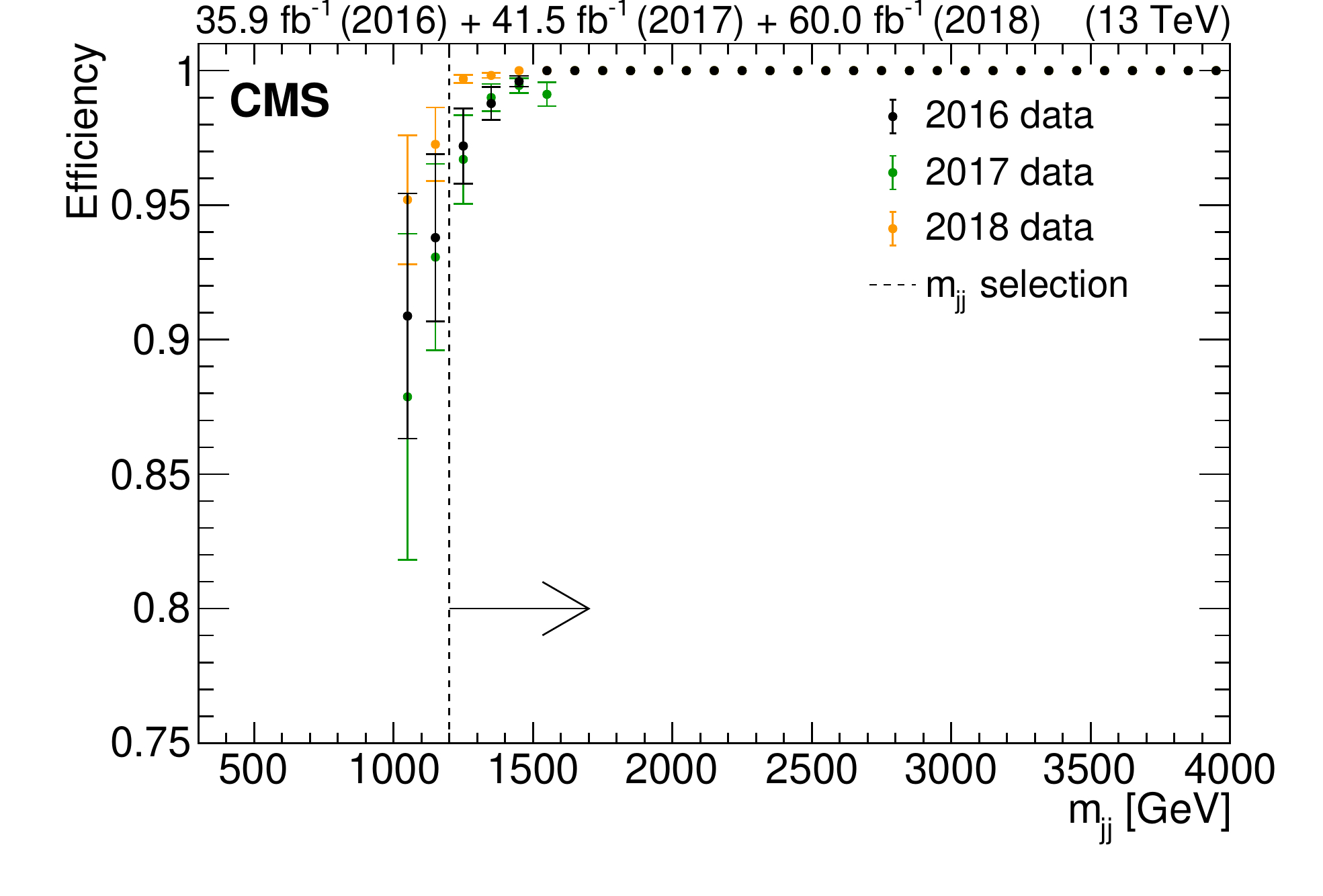}
        \caption{
        The efficiency of the full trigger selection as a function of $\mjj$,
        shown separately for 2016, 2017, and 2018 data.
        The minimum $\mjj$ considered in the analysis is 1200\GeV and
        is marked with a dashed line and an arrow. The efficiency below $\mjj$ of 1000\GeV
        is not measured. The points for 2017 and 2018 data are not visible in the plateau because
        they are overlapped by the points for 2016 data.
        }
        \label{figs:trigger}
\end{figure}

Figure~\ref{figs:trigger} shows the trigger efficiency derived from
2016, 2017, and 2018 data separately.
Simulated samples are corrected using the efficiency measurement from the corresponding
data-taking year.

A trigger inefficiency referred to as ``prefiring'' occurred during the
2016 and 2017 data taking. Over that time period, a gradual shift in the timing of triggering systems
based on the ECAL in the endcap caused certain events to not be recorded. Event corrections
were calculated from data and applied to the 2016 and 2017 simulations to model this inefficiency.
The uncertainties in these corrections are taken as systematic uncertainties.

The SM \ttbar and single top quark Monte Carlo (MC) simulated samples are used as
templates for background estimation in the maximum likelihood fit to data.
A scale factor is applied to the generated top quark \pt
spectrum to correct for the differences between data and
\ttbar simulation.
It is based on a dedicated measurement~\cite{CMS-PAS-TOP-16-011,CMS-PAS-TOP-16-008}, in which
the ratio of the distribution of the top quark \pt measured in data to the distribution as measured in $\POWHEG$+$\PYTHIA$ is derived.
This scale factor may be described by the expression
\begin{linenomath}
    \begin{equation}
        w_{\PQt}(\pt) = \re^{c_{1} 0.0615- c_{2} (0.0005/\GeV) \pt},
        \label{eqn:ptrw}
    \end{equation}
\end{linenomath}
where $c_{1}$ and $c_{2}$ are taken to be 1, as obtained in Refs.~\cite{CMS-PAS-TOP-16-011,CMS-PAS-TOP-16-008}.
The \pt-dependent event weight is given by $\sqrt{\smash[b]{w_{\PQt}(\pt)w_{\cPaqt}(\pt)}}$, where
$w_{\PQt}(\pt)$ and $w_{\cPaqt}(\pt)$ are evaluated using the top quark and antiquark \pt, respectively.
We use the same form for the scale factor but treat $c_{1}$ and
$c_{2}$ as fit parameters initialized to 1 and constrained in the fit to a 
Gaussian with a width of 0.5.

To simulate the SM \ttbar and single top quark production, we use the $\POWHEG$ v2
~\cite{Re:2010bp,Nason:2004rx,Frixione:2007vw,Alioli:2010xd,Powhegttbar} matrix element event generator.
For QCD multijet simulation, we use $\MGvATNLO$ version 5~\cite{Alwall:2014hca} with subversion 2.2.2 for 2016
and 2.4.2 for 2017 and 2018. The QCD multijet simulated samples are
used to derive a scale factor to the multijet background estimation procedure and for cross
checks of self consistency of the background estimate. The \bst signal samples are simulated using $\MGvATNLO$ version 5
over a mass range of 1.4 to 4.0\TeV in steps of 200\GeV.
Subversion 2.2.2 is used for 2016 \bst signal samples with \bst quark masses from 1.4 to 3.0\TeV and 2016 $\bpr$+\PQt
and $\bpr$+\PQb signal samples.
Subversion 2.4.2 is used for 2017 and 2018 \bst signal samples
with \bst quark masses from 1.4 to 3.0\TeV. Subversion 2.6.5 is used for all \bst signal samples with \bst quark
masses above 3.0\TeV. For the $\bpr$ signal simulations, we use
samples based on the 2016 conditions and scale the final distributions to the integrated luminosity of the full data set,
after correcting for differences in annual selection efficiencies that are measured with the $\bst$ signal samples.
We simulate $\bpr$ quark masses from 1.4 to 1.8\TeV in steps of 100\GeV.

Hadronization and parton showering are simulated using the $\PYTHIA$ 8 software package ~\cite{Sjostrand:2014zea}.
The NNPDF3.0~\cite{Ball_2015} parton distribution functions (PDFs)
are used with the CUETP8M1~\cite{CUETP8M1} underlying event tune for the 2016 simulations
and the NNPDF3.1~\cite{Ball_2017} PDFs are used with
the CP5~\cite{CP5} underlying event tune for the 2017 and 2018 simulations. The CMS detector simulation is performed
with $\GEANTfour$~\cite{GEANT4}.
Pythia version 8.212 is used for all the 2016 simulations with the exception
of 2016 \bst signal samples with \bst quark masses from 1.4 to 3.0\TeV, which use version 8.226, and 2016 $\bpr$
signal samples, which use version 8.205.
Pythia version 8.230 is used for all the 2017 and 2018 simulations.

To simulate the effect of additional $\Pp\Pp$ collision data
within the same	or adjacent bunch crossings (pileup), additional inelastic events
are superimposed using $\PYTHIA$.  Simulated samples are then reweighted to correct
the pileup simulation, using the total inelastic cross section
of $69\unit{mb}$~\cite{Aaboud:2016mmw,Sirunyan:2018nqx}
to estimate the distribution of the number of primary vertices in data.

\section{Event reconstruction}
\label{sec:reconstruction}

The candidate vertex with the largest value of summed physics-object
$\pt^2$ is taken to be the primary $\Pp\Pp$ interaction vertex. The physics
objects are the jets, clustered using the anti-$\kt$ jet finding
algorithm~\cite{Cacciari:2008gp,Cacciari:2011ma} with a distance parameter of $R=0.4$
and with the tracks assigned
to candidate vertices as inputs, and the associated missing transverse momentum,
taken as the negative vector sum of the transverse momentum of those jets.

A particle-flow algorithm~\cite{CMS-PRF-14-001} aims to reconstruct and
identify each individual particle in an event, with an optimized combination
of information from the various elements of the CMS detector.
The energy of muons is obtained from the curvature of the corresponding track.
The energy of charged hadrons is determined from a combination of their
momentum measured in the tracking detector and the matching ECAL and HCAL energy deposits,
corrected for the response function of the calorimeters to hadronic showers.
Finally, the energy of neutral hadrons is obtained from the corresponding
corrected ECAL and HCAL energies. Jets are clustered with the anti-$\kt$ jet finding algorithm,
using all particle-flow objects as input. 
Jet momentum is determined as the vectorial sum of all particle momenta
in the jet.
Jets with a distance parameter of $R = 0.8$
are used to reconstruct the top jet and \PW jet candidates in an event.

The pileup per particle identification (PUPPI) algorithm~\cite{Bertolini:2014bba}
is used to mitigate the effect of pileup at the reconstructed particle level,
making use of local shape information, event pileup properties, and tracking
information. A local shape variable is defined, which distinguishes between
collinear and soft diffuse distributions of other particles surrounding the
particle under consideration. The former is attributed to particles originating
from the hard scatter and the latter to particles originating from pileup interactions.
Charged particles originating from pileup
vertices are discarded. For each neutral particle, a local shape variable
is computed using the surrounding charged particles compatible with the
primary vertex within the tracking detector acceptance ($\abs{\eta} < 2.5$), and using
both charged and neutral particles in the region outside of the tracking detector
coverage. The momenta of the neutral particles are then rescaled according
to the probability that they originate from the primary interaction vertex as
deduced from the local shape variable, superseding the need for jet-based
pileup corrections~\cite{Sirunyan:2020foa}.

Jet energy corrections are derived from simulation studies so that the
average measured response of jets becomes identical to that of the jets from the reconstructed
particle level. In situ measurements of the momentum balance in dijet, photon+jet,
Z+jet, and multijet events are used to determine any residual differences
between the jet energy scale in data and in simulation, and appropriate
corrections are made~\cite{Khachatryan:2016kdb}. Additional selection criteria
are applied to each jet to remove jets potentially dominated by instrumental
effects or reconstruction failures~\cite{CMS-PAS-JME-16-003}.

\subsection{Top quark identification}
\label{sec:toptagging}
    The soft-drop algorithm~\cite{Larkoski:2014wba}, a generalization of the
    modified mass drop algorithm~\cite{Dasgupta:2013ihk,Butterworth:2008iy},
    with angular exponent $\beta = 0$ and soft threshold $z = 0.1$, is applied to all jets in the event
    to reconstruct the jet mass and to identify subjets, and includes a grooming step to remove soft radiation,
    including pileup. We only consider top jets with a minimum soft-drop mass of 65\GeV.

    The $N$-subjettiness algorithm~\cite{Thaler:2010tr} defines $\tau_{N}$ variables,
    which describe the consistency between the jet energy deposits and the number of assumed
    subjets, $N$.  When compared to jets originating from a gluon or a light quark, a top
    jet is more consistent with three hard decay products, and the ratio of $\tau_3$
    and $\tau_2$ allows top jets to be distinguished from QCD multijet background~\cite{Thaler_2012}.
    A lower ratio indicates the jet is more consistent with a three-pronged structure
    than a two-pronged structure.

    The \bst signal region selection requires $\tau_3/\tau_2 < 0.65$.
    The $N$-subjettiness ratios are correlated with the jet mass, so a relatively
    loose selection for the signal region is used to avoid biasing the mass distribution of multijet processes.

    We also require the top jet to contain a subjet from the soft-drop algorithm to be identified 
    as a bottom quark by the DeepCSV algorithm~\cite{Sirunyan:2017ezt}. The combination of the
    $\tau_3/\tau_2$ and DeepCSV selections has a QCD jet misidentification rate of approximately 1\% and a
    top tag signal efficiency of approximately 45\%~\cite{CMS-PAS-JME-16-003,CMS-PAS-JME-18-002}. This selection has been chosen because it leads to an optimal
    sensitivity of the cross section limits.

    A jet that passes both the $\tau_3/\tau_2$ and
    DeepCSV $\PQb$ tagging selection is considered ``top tagged''. A \pt-dependent correction is
    applied to correct for differences in the top tagging efficiency between data and simulation~\cite{CMS-PAS-JME-18-002}.
    Separate corrections are used based upon the merging of the top quark decay products in simulation. 
    Taking the line defined by the top quark's trajectory as the central axis, three scenarios are considered.
    In the first, the three decay products are within $R < 0.8$ of the central axis and the jet is considered ``merged''.
    In the second, two out of three decay products are within $R < 0.8$ of the central axis and the jet is ``semi-merged''.
    Finally, with any other configuration of the three decay products the jet is ``not merged''.
    The merged component is the dominant contribution for the $\bst$ signal process among these three scenarios.

\subsection{\texorpdfstring{\PW}{W} boson identification}
\label{sec:wtagging}
    Similar to top tagging, the \PW boson identification algorithm requires a
    selection based on $\tau_{N}$ and soft-drop mass. The \PW jet is required
    to have a soft-drop mass between 65
    and 105\GeV to be consistent with the \PW boson mass~\cite{PDG}.  The
    ratio of $N$-subjettiness $\tau_{2}$ and $\tau_{1}$ variables is used to select the
    characteristic two-prong structure of a hadronic \PW boson decay
    since the \PW jet
    is more consistent with having two subjets than one. The \bst signal region
    selection requires $\tau_2/\tau_1 < 0.4$ for 2016 data and simulation, and
    $\tau_2/\tau_1 < 0.45$ for 2017 and 2018 data and simulation.
    The combined selection on the mass and $\tau_2/\tau_1$ has a QCD jet misidentification rate
    of approximately 10\% and a \PW tag signal efficiency of approximately 80\%,
    which are consistent across the three years~\cite{CMS-PAS-JME-16-003,CMS-PAS-JME-18-002}. This selection was chosen because
    it leads to an optimal sensitivity of the cross section limits.

    A jet that passes
    the $\tau_2/\tau_1$ and soft-drop mass selections is considered ``\PW tagged''.
    Differences in the
    \PW tagging efficiency between data and simulation are corrected using
    simulation-to-data weights~\cite{CMS-PAS-JME-18-002}.
    Additionally, differences in the soft-drop mass scale and resolution
    between data and simulation are accounted for by scaling and smearing the
    soft-drop mass in simulation~\cite{CMS-PAS-JME-16-003}.

\section{Event selection}
\label{sec:selection}
To select signal-like events, two jets are required with $\pt > 400\GeV$
and $\abs{\eta} < 2.4$. Only the two jets with the highest \pt are considered in the following.
The jets are required to satisfy that the difference in rapidity, $\abs{\Delta y}$,
be less than 1.6 and that $\abs{\Delta \phi}$ be greater than $\pi/2$. The $\abs{\Delta \phi}$ requirement
selects back-to-back dijet events while the $\abs{\Delta y}$ requirement suppresses multijet events
with high $\mtw$, which arise from the scattering of valence quarks.
These requirements comprise the "preselection", with an
event then being selected as signal if one of the two jets is \PW
tagged and the other is top tagged.

Because the background estimate relies on data in a control region defined by inverting
the top tag selection, we first require that one of the two jets can be identified as a
\PW jet. In the case that both jets can be \PW tagged, the jet with lower \pt is taken as the \PW boson candidate
in the event. If neither jet can be \PW tagged, the event is not selected.
The jet that is taken as the \PW boson candidate
is referred to as the initially tagged or first jet
and the other jet is called the remaining or second jet.
If the event is selected, it is categorized in either the signal region or the multijet
control region depending on whether the second jet passes the top tagging requirement.
The final selection efficiency for simulated events is calculated as the number of events
that pass the signal selection divided by the number of events generated.
Over the range of generated \bst quark masses, signals with left-handed couplings are selected with an efficiency of 9--10\%.
Signals with right-handed couplings have slightly higher efficiencies, ranging from 10--11\%,
because of their harder jet \pt spectra.

We additionally define a dedicated \ttbar measurement region. For this,
events are required to pass the preselection but
the \PW tag is changed to a top tag selection for the initial jet tag.
This jet tag also requires a top jet mass value between 105 and 220 GeV to be consistent with the top quark mass~\cite{PDG}.
The second top tag will be used to distribute events between the \ttbar measurement
region selection and the dedicated multijet control region for the \ttbar measurement region.
Additionally, for the initial jet tag, the subjet bottom quark requirement remains the same but a tighter selection of $\tau_3/\tau_2 < 0.54$ is required.
The tighter selection on the initial jet tag increases the relative \ttbar contribution.
The $\tau_3/\tau_2$ selection on the second jet tag remains the same as for the \bst signal region
selection to avoid distorting the
mass distribution because of the correlation described in Section~\ref{sec:toptagging}.
If both jets fulfill the selection of the initial top tag, the jet with the lower \pt is takes as the initially top tagged jet in the event.
The \ttbar background measurement region is described in more detail in Section~\ref{sec:ttbarclosure}.
The four tagging selection regions are summarized in Table~\ref{table:selection}.

\begin{table}
    \centering
    \topcaption{A summary of the four selection regions considered in the likelihood fit to data.
    The columns indicate the possible jet tag for the jet considered in the preselection while
    the rows indicate the possible classification of the second jet when using the top tagging algorithm.}
    \label{table:selection}
    \cmsTable{
        \begin{tabular}{c | c c}
                            & W tag & Top tag \\
            \hline
            Top tag         & Signal region (SR) & \ttbar measurement region ($\ttbar \mathrm{MR}$) \\
            Inverted top tag    & Multijet control region (for SR) & Multijet control region (for $\ttbar \mathrm{MR}$) \\
        \end{tabular}
    }
\end{table}

\begin{figure}[htbp!]
    \centering
    \includegraphics[width=0.45\textwidth]{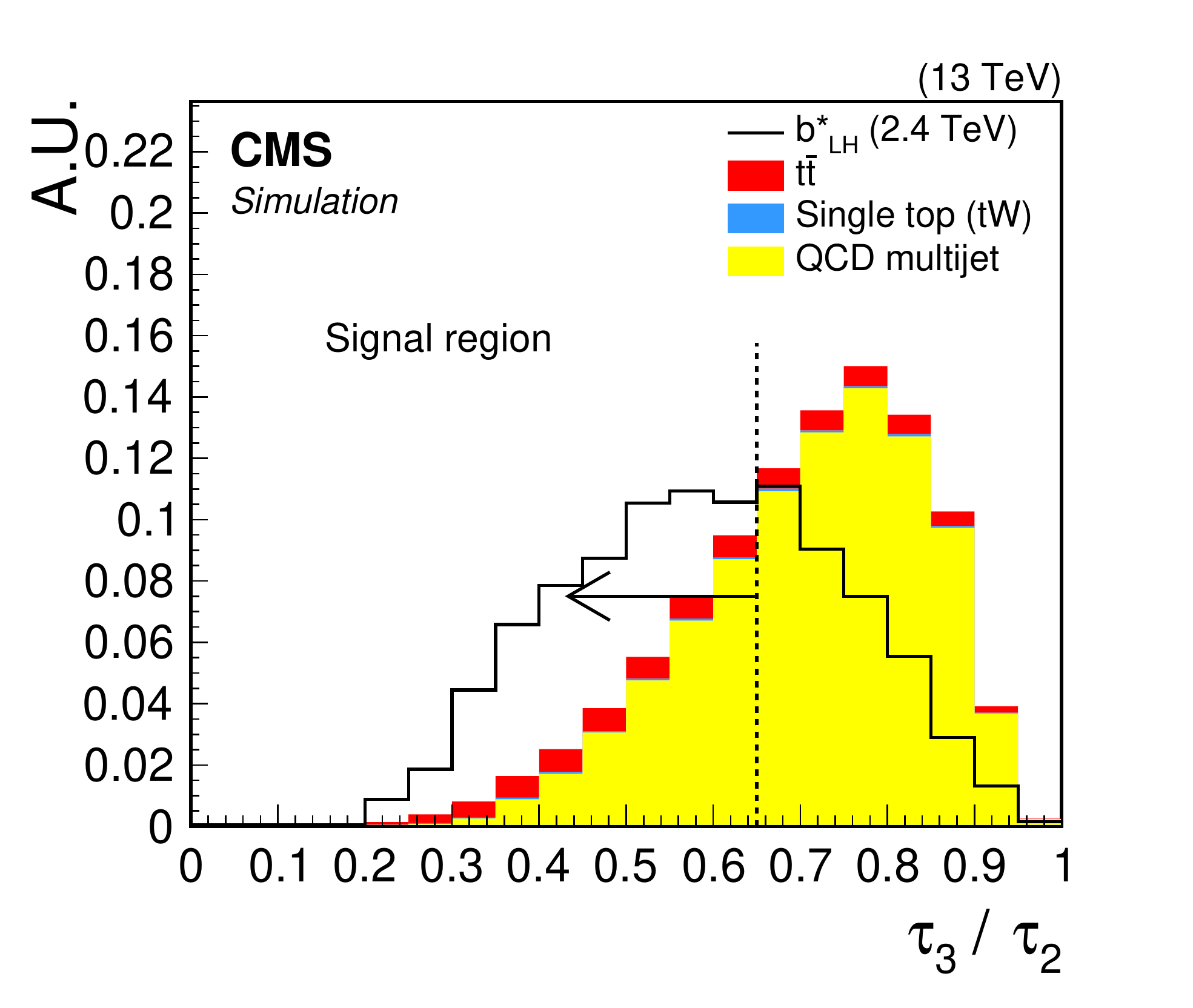}
    \includegraphics[width=0.45\textwidth]{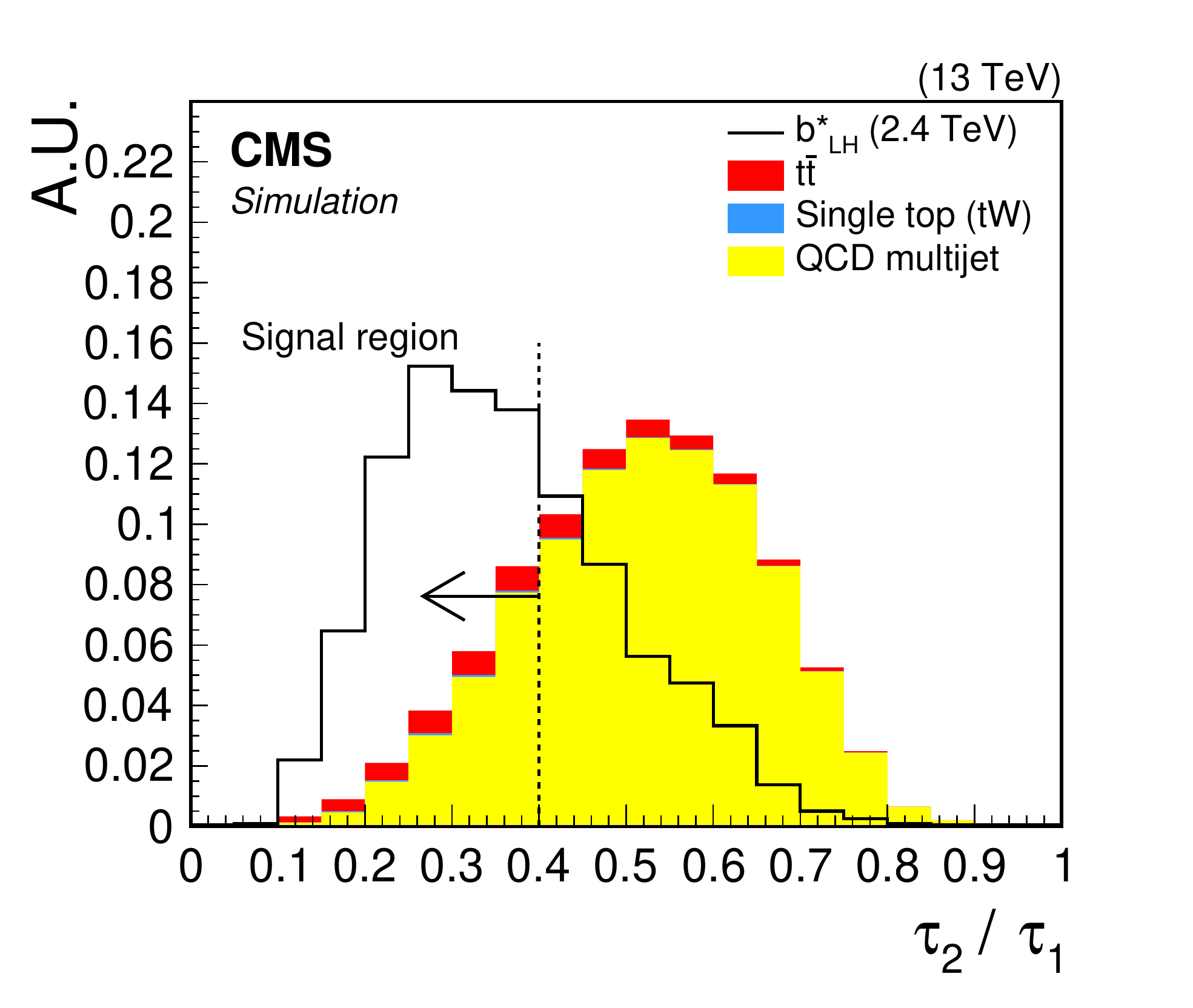}\\
    \includegraphics[width=0.45\textwidth]{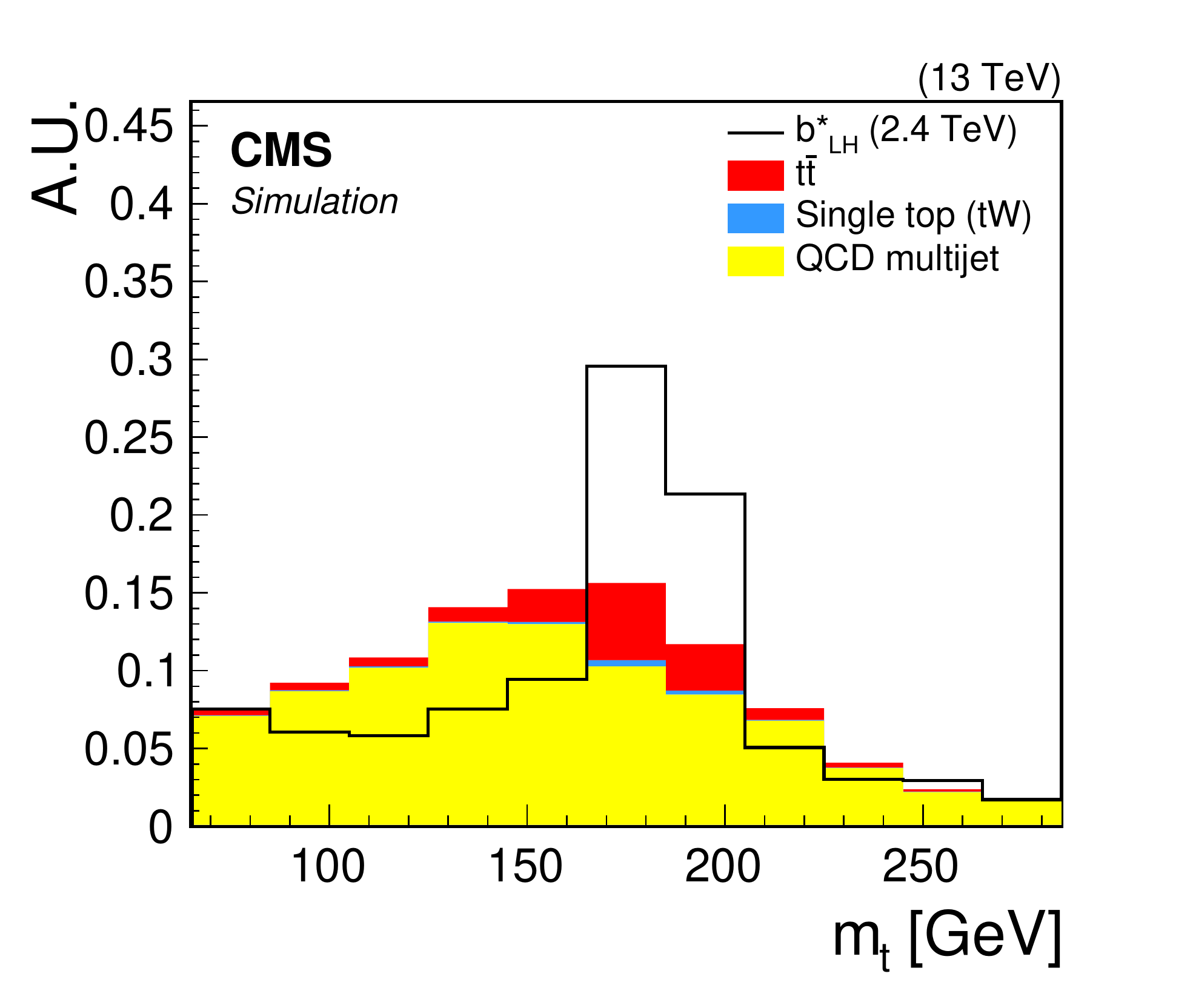}
    \includegraphics[width=0.45\textwidth]{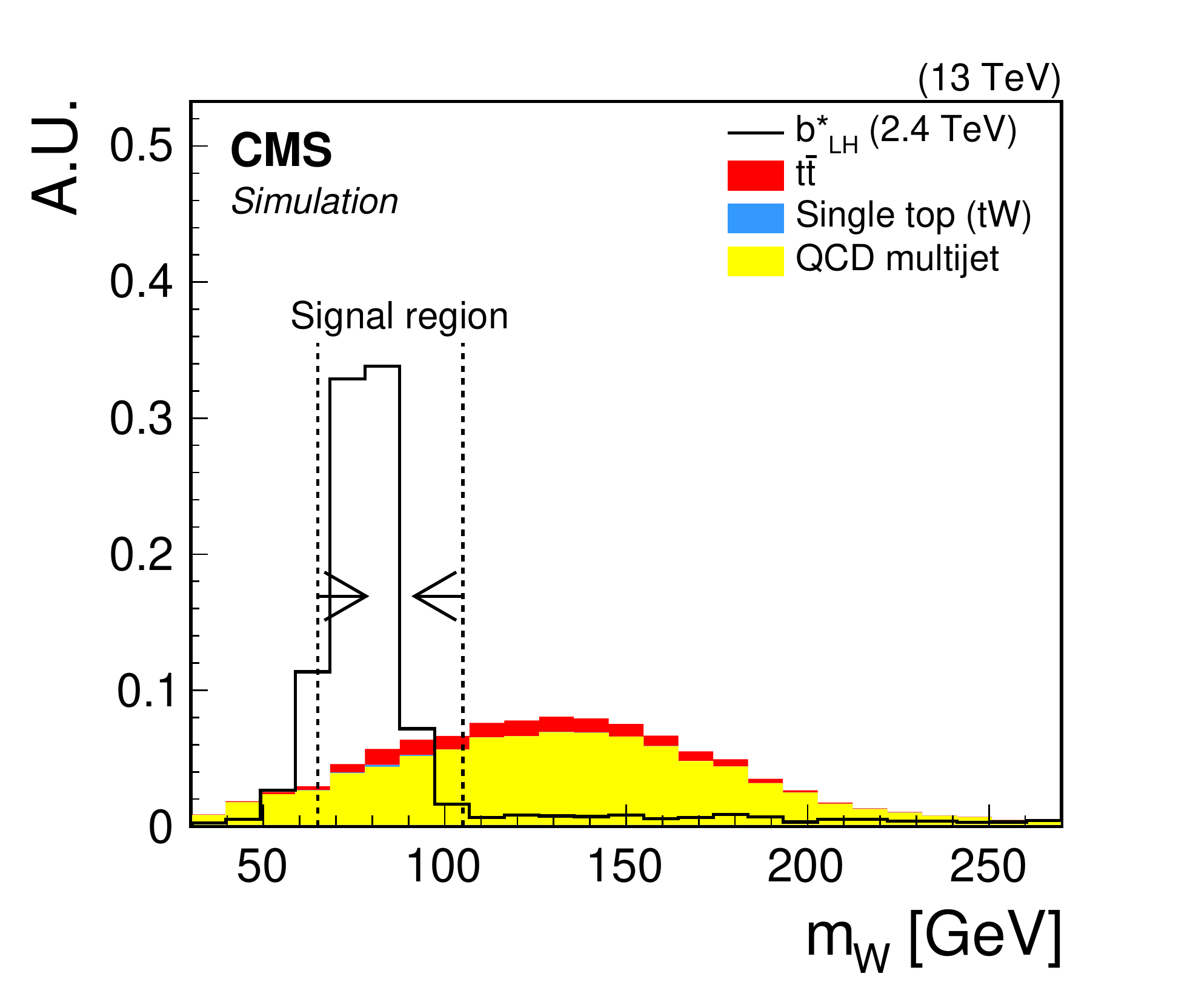}\\
    \includegraphics[width=0.45\textwidth]{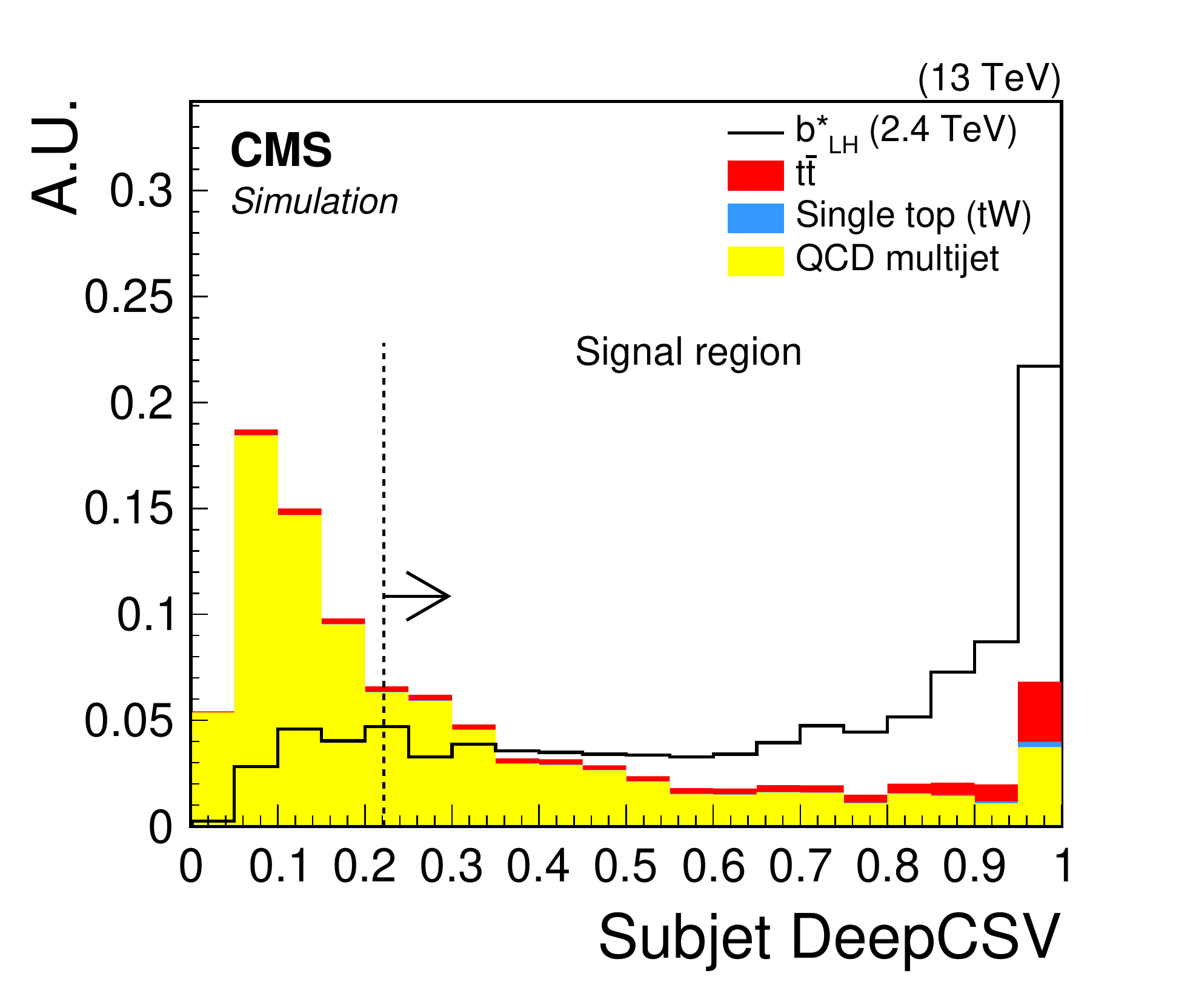}
    \caption{
        The distributions of the discrimination variables used for \PW and top tagging
        for simulation samples. These plots show the top jet $\tau_3/\tau_2$ (upper left),
        the \PW jet $\tau_2/\tau_1$ (upper right), the top tag soft-drop mass (middle left),
        the \PW tag soft-drop mass (middle right), and the subjet \PQb-tagging discriminant (lower).
        The \bst signal sample is represented with the solid line.
        The area of the total background contribution and the area of the signal
        component are separately normalized to unity. All analysis
        selections of the signal region are applied with the exception of the variable being plotted.
        Also shown are vertical dashed lines and arrows, which indicate the
        optimized selection used for events in the signal region. For the \bst signal sample,
        the top tag soft-drop mass spectrum exhibits a resonance in the first bin near the \PW mass, which
        is comprised of \PW jets that have been misidentified as top jets.
    }
    \label{figs:variables}
\end{figure}

Comparisons of the $N$-subjettiness ratio, soft-drop mass, and DeepCSV algorithm score
in simulation between signal and background events are shown in Fig.~\ref{figs:variables}.
The QCD contribution to the multijet background is shown, but the {\PW}+jets and {\PZ}+jets
contributions are omitted since simulations of these processes are not used in this analysis (as discussed in Section~\ref{sec:qcdBackgroundEstimationProcedure}).

\section{Statistical model and background estimation}
\label{sec:background}
The background for this analysis is comprised of multijet, \ttbar, and
$\tw$-channel single top production. The multijet component is estimated
from data while the \ttbar and single top
components are obtained by fitting simulation templates to data.

The $\mt$ range considered is larger than the signal mass window of 105 to 220\GeV 
defined in Section~\ref{sec:selection}. As shown in Fig.~\ref{figs:variables}, an $\mt$ selection is not efficient at discriminating
signal from \ttbar background. However, by using $\mt$ as one of the two measurement dimensions,
one can constrain the multijet background
in the $\mt$ sidebands while distinguishing the multijet background from the top quark backgrounds
in the $\mt$ signal region. 
Thus, the $\mt$ range comprises both the signal peak region and the lower and upper sidebands of the peak.
The signal region considers the range of 65 to 285\GeV while the \ttbar measurement region exists between 105 and 285\GeV,
where the lower mass bound of 105\GeV is used to ensure the orthogonality with the \PW jet mass window of the signal region. 

For each bin in the two-dimensional $(\mt,\mtw)$ distribution, we compare the number of
expected events from both the background-only and signal-plus-background hypotheses with the
number of observed events in data.

The expected number of events from
\bst quark production is calculated as
$N_{\textrm{expected}} = \sigma_{\bst} \mathcal{B}(\bst \to \tw \to \text{hadrons}) \varepsilon L$, where $\sigma_{\bst}$
is the \bst quark cross section,
$\mathcal{B}(\bst \to \tw \to \text{hadrons})$ is the branching fraction of $\bst \to \tw$
in the fully hadronic decay mode,
$\varepsilon$ is the product of the acceptance and the efficiency, and $L$ is
the integrated luminosity of the data set.

A likelihood fit to data is used to test the signal hypothesis, where
the total background model is constructed as a sum of the
individual background contributions using a Poisson model for each bin of the
$(\mt,\mtw)$ distribution.

The number of expected events with failing, $n_{\text{F}}$, and passing, $n_{\text{P}}$,
top tags in a given bin is given by
\begin{linenomath}
    \begin{align}
        n_{\text{F}}(i,\vec{\theta}) &= n_{\text{F}}^{\text{QCD}}(i) + n_{\text{F}}^{\ttbar}(i,\vec{\theta}) + n_{\text{F}}^{\text{single top}}(i,\vec{\theta}) + n_{\text{F}}^{\text{signal}}(i,\vec{\theta})\\
        n_{\text{P}}(i,\vec{\theta}) &= n_{\text{P}}^{\text{QCD}}(i) + n_{\text{P}}^{\ttbar}(i,\vec{\theta}) + n_{\text{P}}^{\text{single top}}(i,\vec{\theta}) + n_{\text{P}}^{\text{signal}}(i,\vec{\theta}),
    \end{align}
\end{linenomath}
where $i$ is a bin in the $(\mt,\mtw)$ plane, and
$\vec{\theta}$ is the set of all nuisance parameters that quantify the systematic uncertainties, as
described in Section~\ref{sec:systematics}.
The variable $n_{\text{F}}^{\text{QCD}}(i)$ is an unconstrained
positive real number. Finally, $n_{\text{P}}^{\text{QCD}}(i)$ is given by
\begin{linenomath}
    \begin{equation}
        n_{\text{P}}^{\text{QCD}}(i) = n_{\text{F}}^{\text{QCD}}(i) f(\mt,\mtw),
        \label{eq:passDef}
    \end{equation}
\end{linenomath}
where $f(\mt,\mtw)$ is a transfer function defined by the ratio of top tagging pass and fail events, and is
described in Section \ref{sec:qcdBackgroundEstimationProcedure}.

The negative log-likelihood is then
\begin{linenomath}
    \begin{equation}\label{eq:nLL}
        \begin{split}
        -\ln L(\vec{d};\vec{\theta}) =
            \sum_{i=1}^{N_{\text{bins,F}}} \left[n_{\text{F}}(i,\vec{\theta}) - d_{\text{F}}(i) \ln n_{\text{F}}(i,\vec{\theta}) \right]
            + \sum_{i=1}^{N_{\text{bins,P}}} \left[n_{\text{P}}(i,\vec{\theta}) - d_{\text{P}}(i) \ln n_{\text{P}}(i,\vec{\theta}) \right],
        \end{split}
    \end{equation}
\end{linenomath}
where $N_{\text{bins,F}}$ and $N_{\text{bins,P}}$ are the total
number of bins and $d_{\text{F}}(i)$ and
$d_{\text{P}}(i)$ are the number of observed events in a given bin, for the
fail and pass distributions, respectively. Thus, there is one likelihood
which combines four separate categories --- signal region ``pass'' and
``fail'' and \ttbar measurement region ``pass'' and ``fail''.

\subsection{Multijet background estimate}
\label{sec:qcdBackgroundEstimationProcedure}

After applying the kinematic selection along
with the \PW jet identification, we define the ratio of the multijet background
distributions that pass and fail the top tagging requirement in data and
QCD multijet MC simulation as $\rpfdata$ and $\rpfmc$, respectively.
Because of the combinatorial nature of multijet processes,
$\rpfdata$ and $\rpfmc$ are both smooth as a function of $\mt$ and $\mtw$.
The data-to-simulation ratio of these ratios is therefore also
smooth and can be used to correct for differences in simulation and data
by parameterizing it with an analytic function, $\rrat$.

While $\rpfdata$ could also be described by analytic functions,
isolated features of the shape can be factored out by using simulation. By
factoring out $\rpfmc$, the fit of the analytic function to data is only
responsible for describing the residual differences between data and
simulation that can be parameterized with fewer parameters than
the shape of $\rpfdata$.

The number of events in a given bin of the passing category can then
be estimated from the equation
\begin{linenomath}
    \begin{equation}
        n_{\text{P}}^{\text{QCD}}(i) = n_{\text{F}}^{\text{QCD}}(i) \rpfmc \rrat,
        \label{eq:passDefTF}
    \end{equation}
\end{linenomath}
where $f(\mt,\mtw)$ has been replaced by $\rpfmc \rrat$ and $\rrat$ is a surface
parameterized by the product of two one-dimensional polynomials in the $(\mt,\mtw)$ plane with
coefficients determined from the fit to data.
A second-order polynomial was chosen for the $\mt$ axis and a first-order polynomial was chosen for the
$\mtw$ axis. These choices were based on a Fisher test~\cite{10.2307/2340521} where
polynomial terms were added until the $p$-value obtained in the test
was less than 0.95. The parameters of the
two-dimensional polynomials are uncorrelated between years. The form of $\rrat$ is then
\begin{linenomath}
    \begin{equation}
        (p_0+p_1 \mt+p_2 \mt^2)(1+p_3 \mtw).
    \end{equation}
\end{linenomath}
To reduce the effect of statistical fluctuations when calculating $\rpfmc$ in the QCD multijet
simulation, the pass and fail distributions are smoothed by using an adaptive kernel density
estimate~\cite{Cranmer2000KernelEI} (KDE) prior to calculating the ratio.
Additionally, the residual contributions from the {\PW}+jets and {\PZ}+jets backgrounds are
accounted for in this analysis, as they are absorbed by the unconstrained $\rrat$ function.

\subsection{Top quark measurement region}
\label{sec:ttbarclosure}
By performing the maximum likelihood fit to data in the signal region
simultaneously with the \ttbar background enriched measurement region, we further constrain
the \ttbar contribution to the total background estimate.
In particular, this region is used to make measurements of the $c_{1}$
and $c_{2}$ fit parameters of Eq.~(\ref{eqn:ptrw}).

The \ttbar measurement region is evaluated in the $(\mt,\mtt)$ plane,
where $\mt$ is the mass of the second jet when using the top tagging
algorithm and $\mtt$ is the invariant mass of the \ttbar pair.
Only the multijet and \ttbar SM processes are considered in this
selection since the single top quark contribution is negligible.

The strategy to estimate the multijet background in the \ttbar measurement
region is similar to the signal region.
The $\rrattt$ in this region is parameterized with the same polynomial form as in the signal region,
but the parameters are uncorrelated with those of the signal region. Additionally, the $\rpfmctt$
is derived using QCD multijet simulation events that pass the same
selection as the \ttbar measurement region. Events from {\PW}+jets and {\PZ}+jets backgrounds
are suppressed by the initial top tag requirement, and any that remain are accounted for by the multijet
background model as they are in the signal region.

The negative log-likelihood calculated from the \ttbar measurement region is constructed similarly to
Eq.~(\ref{eq:nLL}). The total negative log-likelihood is obtained from the sum of the negative log-likelihoods
of the signal region and the \ttbar measurement region.
Because the fit to data can constrain the \ttbar background in both
selections, the values of the free parameters that determine the shape
and normalization of the \ttbar simulation are constrained by the
simultaneous fit to the \ttbar- and signal-enriched selections.

\section{Systematic uncertainties}
\label{sec:systematics}
This analysis takes into account several systematic uncertainties that can affect
both the shape and normalization of the simulation.

Normalization uncertainties include those in the production cross
section and in the measured integrated luminosity of the data.
The uncertainties in the \ttbar and single top $\tw$-channel production are
taken as 20 and 30\%, respectively, to account for
the uncertainties in the cross section and in the factorization and renormalization scales of each process.
Specifically, these values were chosen based on the largest variations in
yield of the simulated samples from varying the factorization and renormalization scales.
The uncertainty in the measured integrated luminosity is 1.8\%~\cite{CMS-PAS-LUM-17-001,CMS-PAS-LUM-17-004,CMS-PAS-LUM-18-002}
for the complete Run 2 (2016--2018).

Several uncertainties exist that affect both the shape and normalization
of the $(\mt,\mtw)$ distributions. The uncertainties in the jet energy scale
and resolution are estimated through variations
in \pt and $\eta$ of the PUPPI jets~\cite{Khachatryan:2016kdb}.
The uncertainty in the pileup reweighting correction is evaluated by varying
the total inelastic cross section by $\pm$4.6\%~\cite{Aaboud:2016mmw}.
The uncertainty in the trigger correction is taken into account as a variation of
one half of the trigger inefficiency.
The uncertainty in the PDFs is
derived by either evaluating the root-mean-square of the set of NNPDF MC replicas or by evaluating the
contributions of eigenvectors provided in a Hessian set~\cite{Butterworth_2016},
depending on whether the PDF set represents variations as MC replicas or Hessian eigenvectors.
The uncertainty due to differences in the data and simulation efficiency
for the top jet tagging algorithm is evaluated by using the
variations of the top tagging scale factor~\cite{CMS-PAS-JME-18-002}.
The scale factors and uncertainties vary depending on the merging scenarios defined in Section~\ref{sec:toptagging}.
The \PW tagging uncertainty
is evaluated from variations in the \PW tagging scale factor and includes
an additional uncertainty when extrapolating
to jets outside of the \pt region used to extract the scale factor.
Additionally, the uncertainty in the \PW tagging soft-drop mass selection is
evaluated from variations in the jet mass scale and resolution~\cite{CMS-PAS-JME-18-002}.
No variations in the jet mass scale and resolution are considered for the candidate top jet
since the effect is negligible with respect to the current results.

Unique to the \ttbar simulation is the uncertainty in the top quark \pt reweighting
procedure described in Section~\ref{sec:dataset}, which is extrapolated to high \pt.
The uncertainty	is represented as uncorrelated variations of $\pm50\%$ in each of the $c_{1}$ and $c_{2}$
parameters from Eq.~(\ref{eqn:ptrw}).

Each uncertainty affecting both the shape and normalization is Gaussian constrained
where the $\pm 1$ standard deviation of each distribution is mapped to the $\pm 1$
standard deviation of the corresponding unit Gaussian constraint.

The uncertainty in the multijet background estimation is taken
from the maximum likelihood fit to data. The parameters of each
two-dimensional polynomial are uncorrelated and fitted freely with no \textit{a-priori} constraints.
An additional uncertainty in the ``bandwidth'' parameter of the KDE
algorithm is accounted for by varying the parameter up and down by 1,
where the nominal value is 4. This
parameter acts as a scale to determine the width of the adaptive kernels.

All systematic uncertainties are considered in the simultaneous fit to data
such that all correlations are preserved. The uncertainties are
always correlated across $\tw$ and \ttbar selections within a given year of data and simulation.
The cross section, PDF, and
top quark \pt reweighting $c_{1}$ and $c_{2}$ uncertainties are individually correlated
across the data-taking years. Table~\ref{table:sys}
summarizes the sources of uncertainty and indicates where correlations between samples exist.

Additionally, Table~\ref{table:sys} includes a calculation of the ``impact'' of a parameter on the
measurement of the final signal strength for a 2.4\TeV \bst quark signal.
This value is calculated by comparing the measured signal strength in the full fit against
the measured signal strength in a fit where the given nuisance parameter has been changed
either ``up'' or ``down'' one standard deviation from its post-fit value in the full fit.

As can be seen in Table~\ref{table:sys}, the multijet estimate from data is the dominant source of background uncertainty
in the measurement of the signal strength. In particular, variations of one post-fit
standard deviation of the linear term in the $\mtw$ axis of the signal region
can change the measurement of the signal strength by approximately 19\%.

\begin{table}[htb]
\centering
\topcaption{Sources of uncertainty that are taken into account in the statistical analysis of the data.
The sources affecting the normalization are given with their percentage uncertainties, while
the sources affecting the shape are listed as ``Shape'' together with the dependent parameter.
The rightmost column indicates the impact of the parameter on the 2.4\TeV
\bst signal strength when the parameter is changed ``up'' and ``down''
by one standard deviation from its post-fit value. For parameters where the uncertainties
are uncorrelated between data-taking years, the average impact is calculated. An impact of $+0.0$ ($-0.0$)
denotes an impact that is less (greater) than 0.1 ($-0.1$) but greater (less) than 0.}
\label{table:sys}
    \cmsTable{
    \begin{tabular}{lcccc}
                            & 							& 								& \multicolumn{2}{c}{Impact} \\
    Source				    & Uncertainty 			    & Samples					    & Up     & Down \\
    \hline
    \ttbar cross section 	& $\pm$20\% 				& \ttbar 						& $-4.6$ & $+4.4\%$\\
    Single top cross section   & $\pm$30\%      			& Single top 					& $+1.2$ & $-1.4\%$\\
    Integrated luminosity    	& $\pm$1.8\% 				& \ttbar, single top, signal  & $+1.6$ & $-1.1\%$\\
    Pileup     				& Shape ($\sigma_{mb}$) 	& \ttbar, single top, signal  & $+0.3$ & $-0.2\%$\\
    Trigger prefiring   	& Shape (\pt, $\eta$)  	& \ttbar, single top, signal  & $+0.0$ & $+0.1\%$\\
    Jet energy scale  			& Shape (\pt)    			& \ttbar, single top, signal  & $+0.3$ & $-0.6\%$\\
    Jet energy resolution 		& Shape (\pt, $\eta$)  	& \ttbar, single top, signal  & $-0.4$ & $-0.5\%$\\
    Jet mass scale  			& Shape ($\mW$)    			& \ttbar, single top, signal  & $-0.1$ & $-0.0\%$\\
    Jet mass resolution 		& Shape ($\mW$)  			& \ttbar, single top, signal  & $+0.0$ & $+0.9\%$\\
    \PW tagging    			& Shape (\pt)    		    & Single top, signal 			& $+0.9$ & $-0.9\%$\\
    \PW tagging: \pt extrapolation & Shape (\pt)    	& Single top, signal            & $+4.9$ & $-4.9\%$\\
    Top tagging, merged		& Shape (\pt)     		& \ttbar, single top, signal  & $+0.2$ & $-0.2\%$\\
    Top tagging, semimerged	& Shape (\pt)     		& \ttbar, single top, signal  & $+1.1$ & $-0.9\%$\\
    Top tagging, not merged	& Shape (\pt)     		& \ttbar, single top, signal  & $-0.1$ & $+0.1\%$\\
    Trigger    				& Shape (\HT)    			& \ttbar, single top, signal  & $+0.3$ & $-0.4\%$\\
    Top quark \pt correction $c_{1}$ 	& Shape (\pt) & \ttbar                      & $-0.3$ & $+0.3\%$\\
    Top quark \pt correction $c_{2}$ 	& Shape (\pt) & \ttbar                      & $-3.9$ & $+3.5\%$\\
    PDF						& Shape $(\mt,\mtw)$		& Signal					    & $+0.1$ & $-0.1\%$\\
    KDE bandwidth				& Shape $(\mt,\mtw)$ 	& Multijet (from simulation)	& $-1.2$ & $+0.2\%$\\
    $\rratsr p_0$     		& Shape $(\mt,\mtw)$  	& Multijet (from data)          & $-4.4$ & $+0.0\%$\\
    $\rratsr p_1$     		& Shape $(\mt,\mtw)$  	& Multijet (from data)          & $-2.0$ & $+2.2\%$\\
    $\rratsr p_2$     		& Shape $(\mt,\mtw)$  	& Multijet (from data)          & $+0.9$ & $-0.8\%$\\
    $\rratsr p_3$     		& Shape $(\mt,\mtw)$  	& Multijet (from data)          & $+18.6$ & $-18.8\%$\\
    $\rrattt p_0$     	& Shape $(\mt,\mtt)$  	& Multijet (from data)          & $-0.4$ & $+0.6\%$\\
    $\rrattt p_1$     	& Shape $(\mt,\mtt)$  	& Multijet (from data)          & $-0.4$ & $+0.6\%$\\
    $\rrattt p_2$     	& Shape $(\mt,\mtt)$  	& Multijet (from data)          & $+0.5$ & $-0.6\%$\\
    $\rrattt p_3$     	& Shape $(\mt,\mtt)$  	& Multijet (from data)          & $-0.6$ & $+0.6\%$\\
    \end{tabular}
    }
\end{table}

\section{Results}
\label{sec:results}

The $(\mt,\mtw)$ and $(\mt,\mtt)$ distributions are used in a simultaneous binned
maximum likelihood fit to data. The signal strength is a free parameter in the model
and the systematic uncertainties are accounted for as nuisance parameters as described in Section~\ref{sec:background}.
Normalization uncertainties are modeled with log-normal priors, and uncertainties affecting
simulation shapes are modeled using a template morphing approach with Gaussian priors.

\begin{figure}[htp!]
    \centering
    \includegraphics[width=0.9\textwidth]{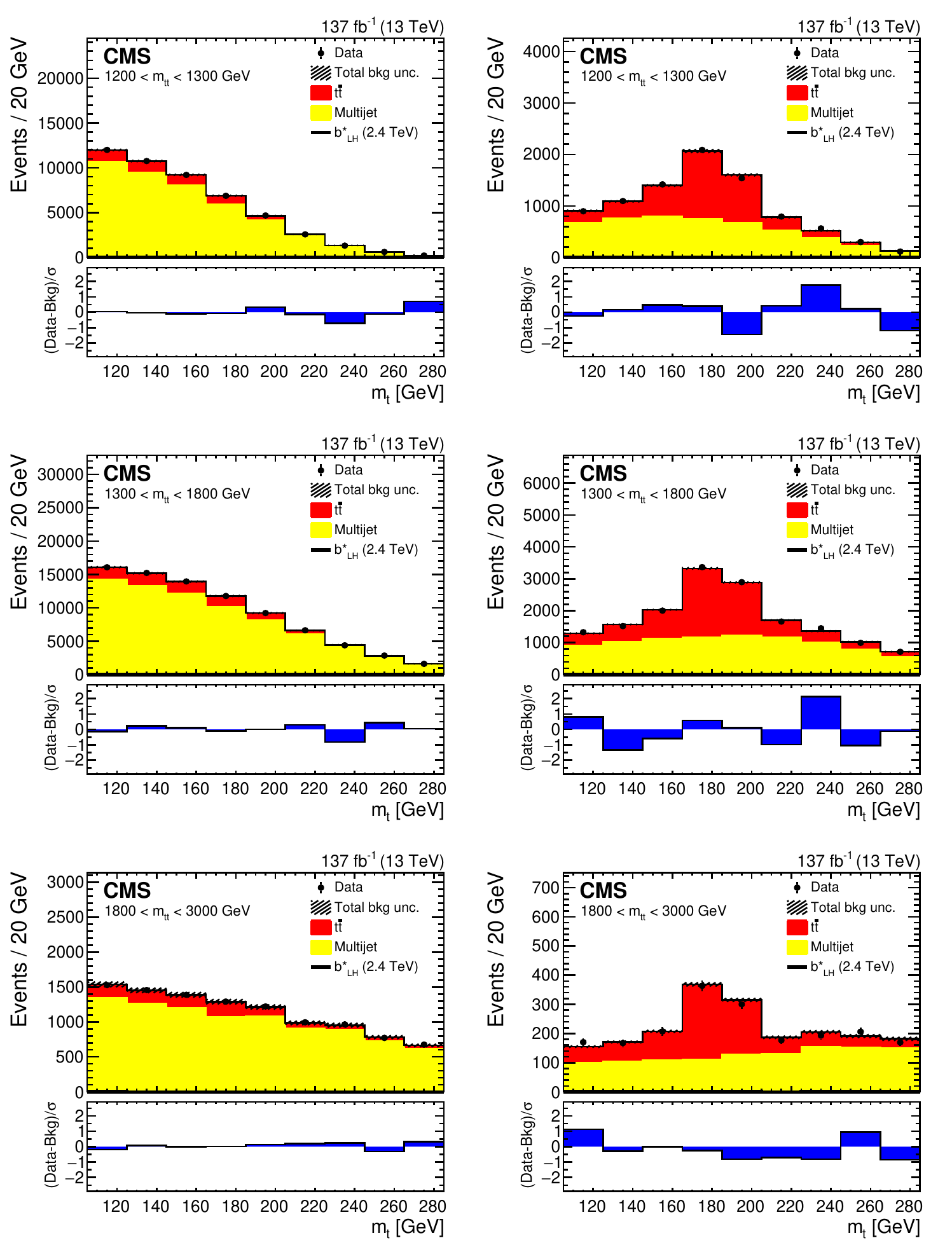}
    \caption{
        Distributions of $\mt$ in the \ttbar measurement region for three intervals of $\mtt$:
        1200--1300\GeV (upper), 1300--1800\GeV (middle), 1800--3000\GeV (lower).
        The data are shown by points with error bars and the individual background contributions by filled histograms.
        The signal is not visible because the contamination in this region is negligible.
        The barely visible shaded region is the
        uncertainty in the total background estimate.
        The left and right columns show distributions for events with the second jet failing and passing the
        top tagging requirement, respectively.
        The lower panels of each figure show the pull, as a function of $\mt$, defined as the difference
        between the number of events observed in the data and
        the predicted background, divided by their combined uncertainty.
    }
    \label{figs:ttRunIIbump}
\end{figure}

\begin{figure}[htp!]
    \centering
    \includegraphics[width=0.9\textwidth]{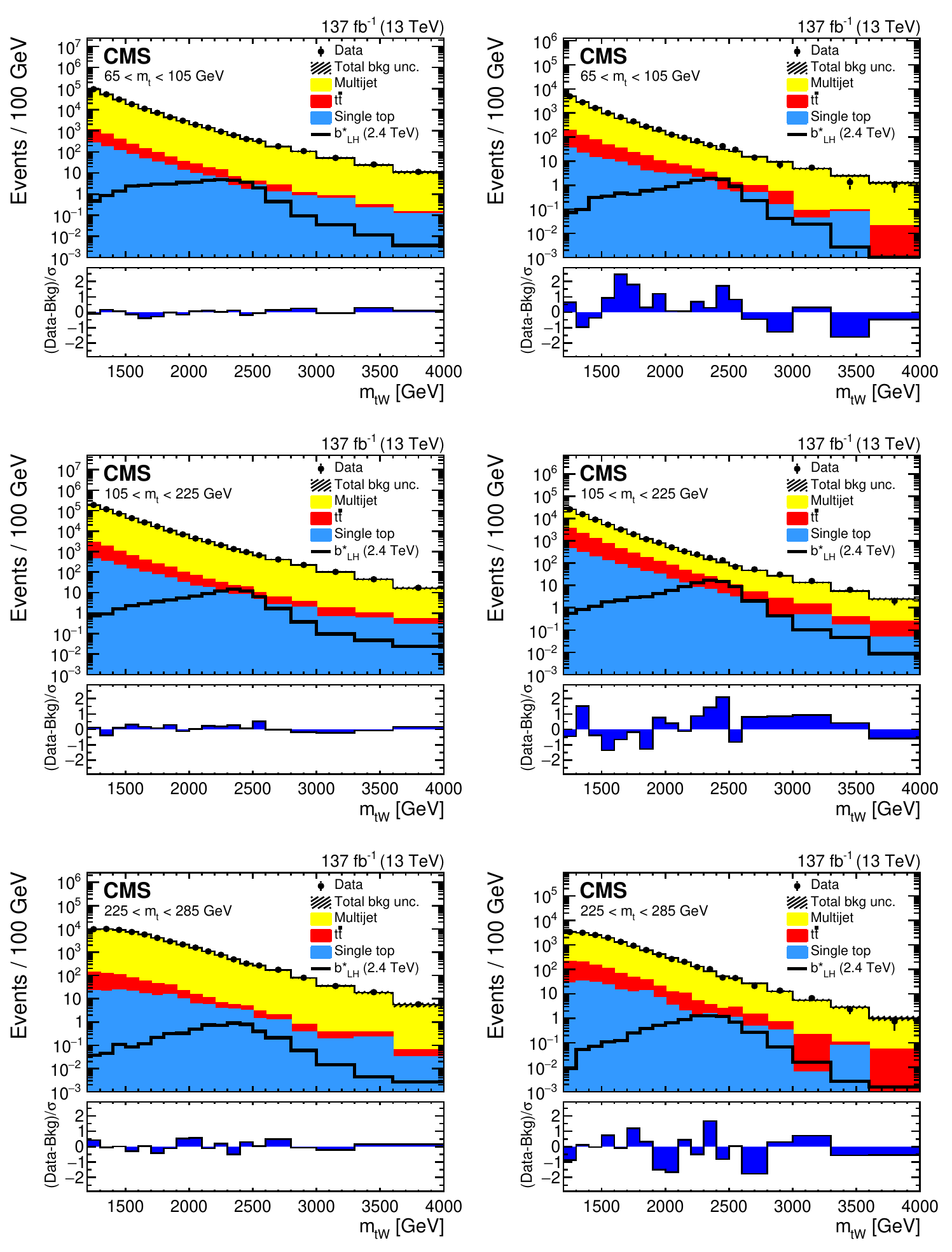}
    \caption{
        Distributions of $\mtw$ in the \bst signal region for three intervals of $\mt$:
        65--105\GeV (upper), 105--225\GeV (middle), and 225--285\GeV (lower).
        The data are shown by points with error bars, the individual background contributions by filled histograms,
        and a 2.4\TeV $\bstl$ signal is shown as a solid line.
        The barely visible shaded region is the
        uncertainty in the total background estimate.
        The left and right columns show distributions for events with a jet failing and passing the
        top tagging requirement, respectively.
        The lower panels of each figure show the pull, as a function of $\mtw$, defined as the difference
        between the number of events observed in the data and
        the predicted background, divided by their combined uncertainty.
    }
    \label{figs:SRRunIIbump}
\end{figure}

While the fit is performed in two dimensions, evaluating the agreement
of the background model with the data is more convenient when examining projections onto one dimension.
The background estimate and measured two-dimensional distributions from the
simultaneous fit of the signal region, \ttbar measurement region, and multijet--enriched control regions
are shown in Figs.~\ref{figs:ttRunIIbump} and~\ref{figs:SRRunIIbump}, respectively,
as one-dimensional projections where either the $\mtt$ or $\mt$ distribution
has been separated into three regions. The lower panels show the pull,
defined as the difference between the number of events observed in the data and
the predicted background, divided by the systematic uncertainty in the background and the
statistical uncertainty in the data, added in quadrature.
All plots shown are for the signal-plus-background hypothesis post-fit,
where the 2.4\TeV $\bstl$ quark sample
is normalized to the post-fit signal cross section.

In Fig.~\ref{figs:ttRunIIbump}, the left column shows
distributions of $\mt$ obtained for the selection of the \ttbar measurement region,
but with a jet failing the top tagging requirement. The right column shows
the same distributions, but for jets passing the top tagging requirement.
The rows give the distributions for separate intervals of $\mtt$.
The background estimation is observed to model the data well in both regions,
validating the estimation of the multijet background and the modeling of the \ttbar simulation.
The contribution from a possible signal is negligible in this region and therefore not visible.

In Fig.~\ref{figs:SRRunIIbump}, distributions of $\mtw$, obtained for events passing the signal region selection are
shown, where the distributions in the left and right columns have been obtained for jets
failing and passing the top tagging requirement, respectively.
Plots in the row are for separate intervals of $\mt$.
The total background estimate agrees with the data within the uncertainties. The largest excess in data relative
to the total background is observed for a left-handed \bst quark with a mass of 2.4\TeV, which
results in a local significance of 2.3 standard deviations.

Additionally, the post-fit top quark \pt reweighting measurements are consistent
with the pre-fit values, and are measured to be $c_{1} = 1.01 \pm 0.25$
and $c_{2} = 1.16 \pm 0.16$.
The agreement of the background-only model is evaluated
using the saturated test statistic~\cite{Baker:1983tu,parametricInference} and has a $p$-value
of 0.3. Additionally, the post-fit nuisance parameter values are consistent with the
pre-fit values and the nuisance parameter values from the background-only model fit are consistent
with those from the signal-plus-background model fit.

Asymptotic frequentist statistics are used to derive exclusion limits on 
$\sigma_{\bst} \mathcal{B}(\bst \to \tw \to \text{hadrons})$ at 95\%
CL~\cite{Cowan_2011}. These limits are derived separately for the $\bstr$, $\bstl$, and $\bstlr$ quark
signal hypotheses. The $\pm$1 and $\pm$2 standard
deviations in the expected limit are derived from pseudo-experiments under the
background-only hypothesis in which the nuisance parameters are randomly varied
within the post-fit constraints of the maximum likelihood fit to data.

\begin{figure}[htbp!]
    \centering
    \includegraphics[width=0.55\textwidth]{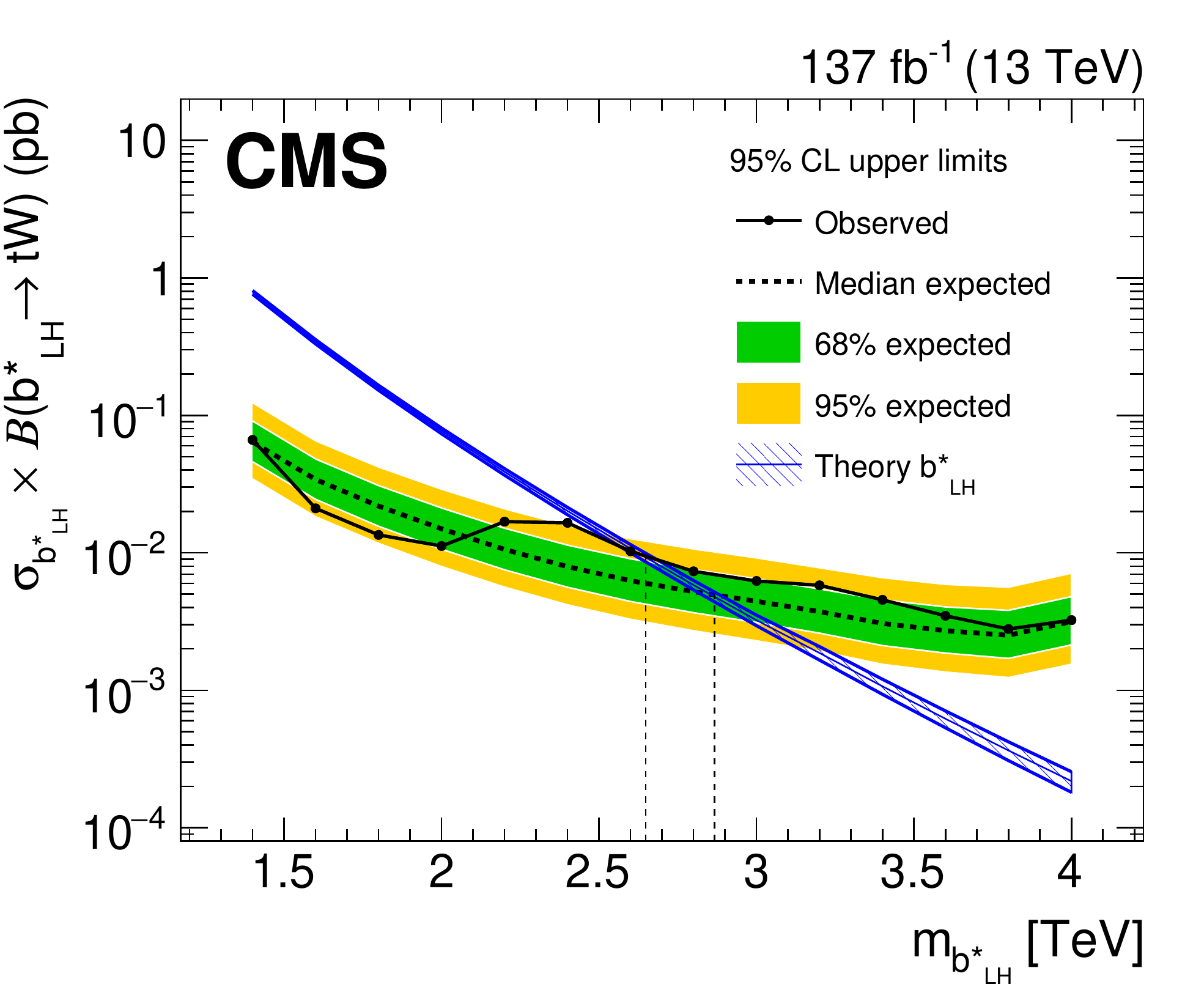}
    \includegraphics[width=0.55\textwidth]{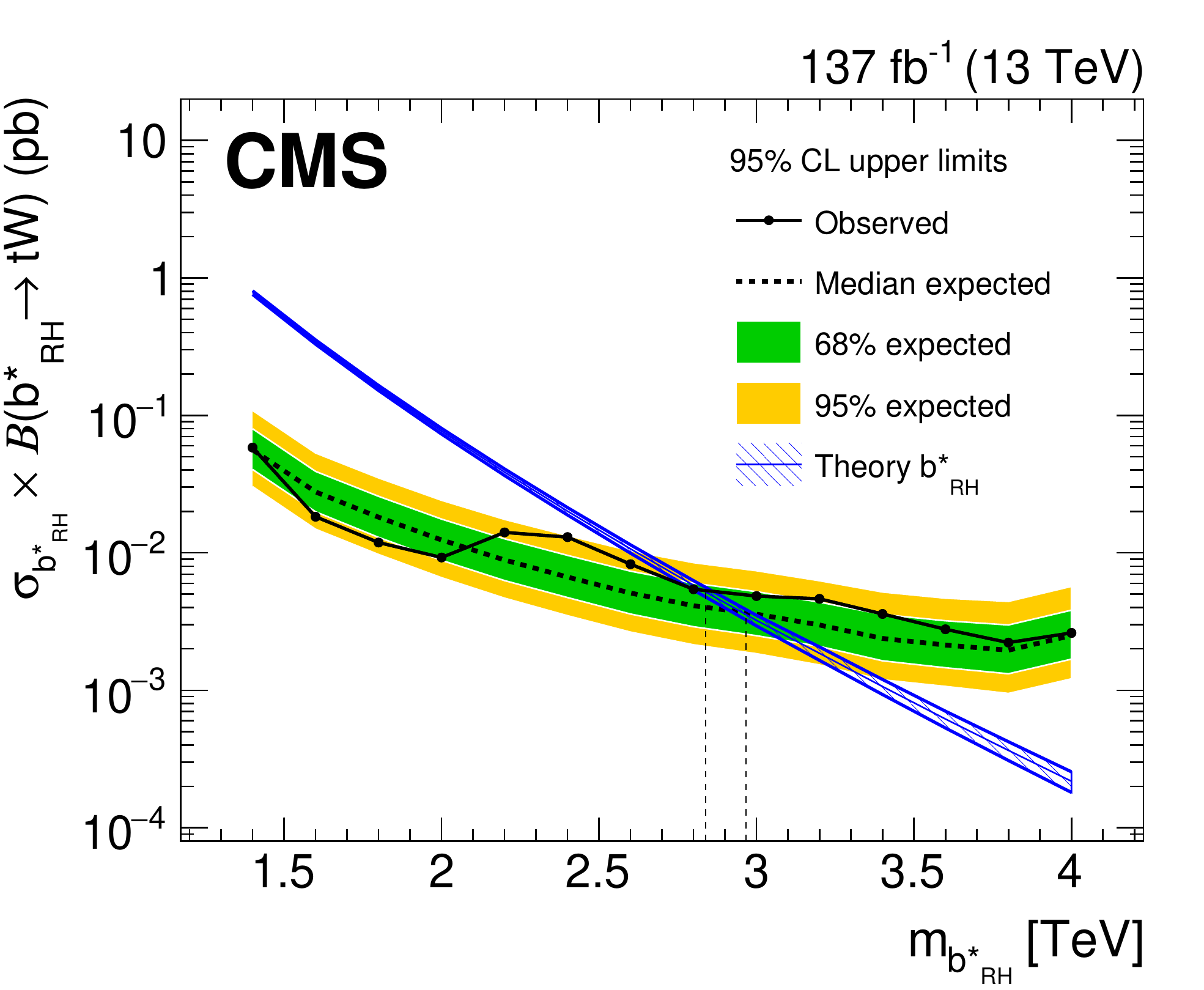}
    \includegraphics[width=0.55\textwidth]{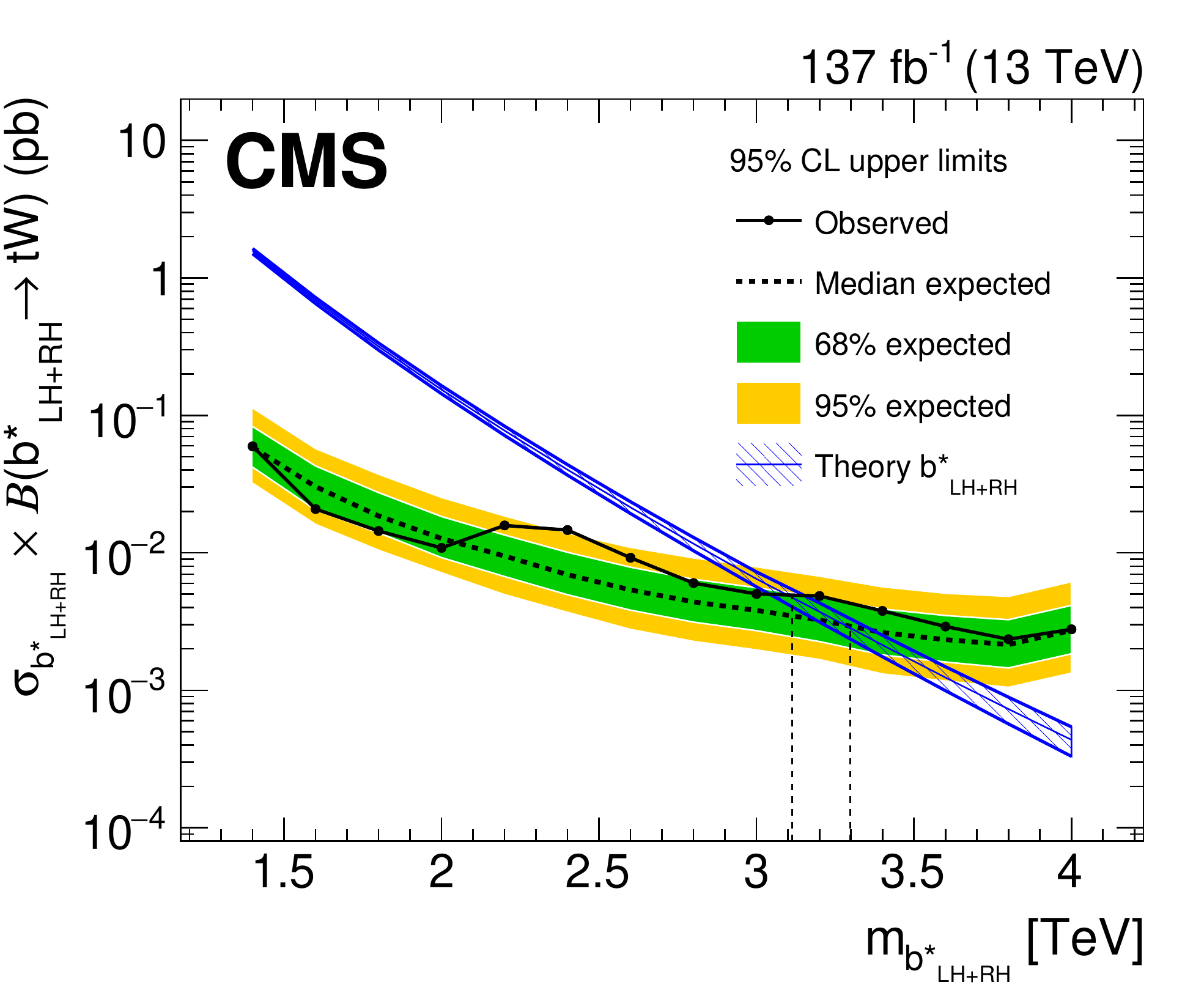}

    \caption{
        Upper limits on the product of the cross section and branching fraction at 95\% $\CL$
        for a $\bstl$ (upper), $\bstr$ (middle), and $\bstlr$ (lower) quark as a function of the \bst quark mass.
        The expected (dashed) and observed (dot-solid) limits, as well as the \bst quark
        theoretical cross sections (shaded-solid), are shown. The vertical dashed lines indicate
        the intersection of the theoretical cross sections with the expected and observed limits.
        The inner and outer shaded areas around the expected limits show the 68\% and 95\% CL intervals, respectively.
        }
    \label{figs:limits}
\end{figure}

The limits are shown in Fig.~\ref{figs:limits}. The theoretical \bst cross sections
included in the figure as a function of \bst quark mass are calculated using $\MGvATNLO$.
Masses below 2.6, 2.8, and 3.1\TeV (2.9, 3.0, and 3.3\TeV) are observed (expected)
to be excluded at 95\% \CL
for the left-handed, right-handed, and vector-like hypotheses, respectively.
These limits nearly doubles the mass exclusions of the previous result~\cite{Khachatryan2016}.

The sensitivity of this analysis can also be compared to the sensitivity of the CMS dijet search~\cite{CMSDijet}.
The branching fraction for $\bst \to \PQb\Pg$ approaches 20\% asymptotically for high masses~\cite{Nutter_2012}.
From the dijet search, the expected upper limit on the product of the cross section and branching fraction for a
resonance decaying to a quark and a gluon is approximately 0.09\unit{pb} at 2\TeV
so the cross section upper limit on \bst quark production is approximately 0.45\unit{pb}.
Using the left-handed couplings result in Fig.~\ref{figs:limits}, this analysis achieves an
expected upper limit of approximately 0.015\unit{pb} at 2\TeV.
With the $\bst \to \tw$ branching fraction of 0.4, the cross section upper limit on $\bstl$ quark production at 2\TeV is
approximately 0.0375\unit{pb}.
Thus, at 2\TeV, this search is about an order of magnitude more sensitive to the excited \bst quark
than the dijet search.

\begin{figure}[htbp!]
    \centering
    \includegraphics[width=0.49\textwidth]{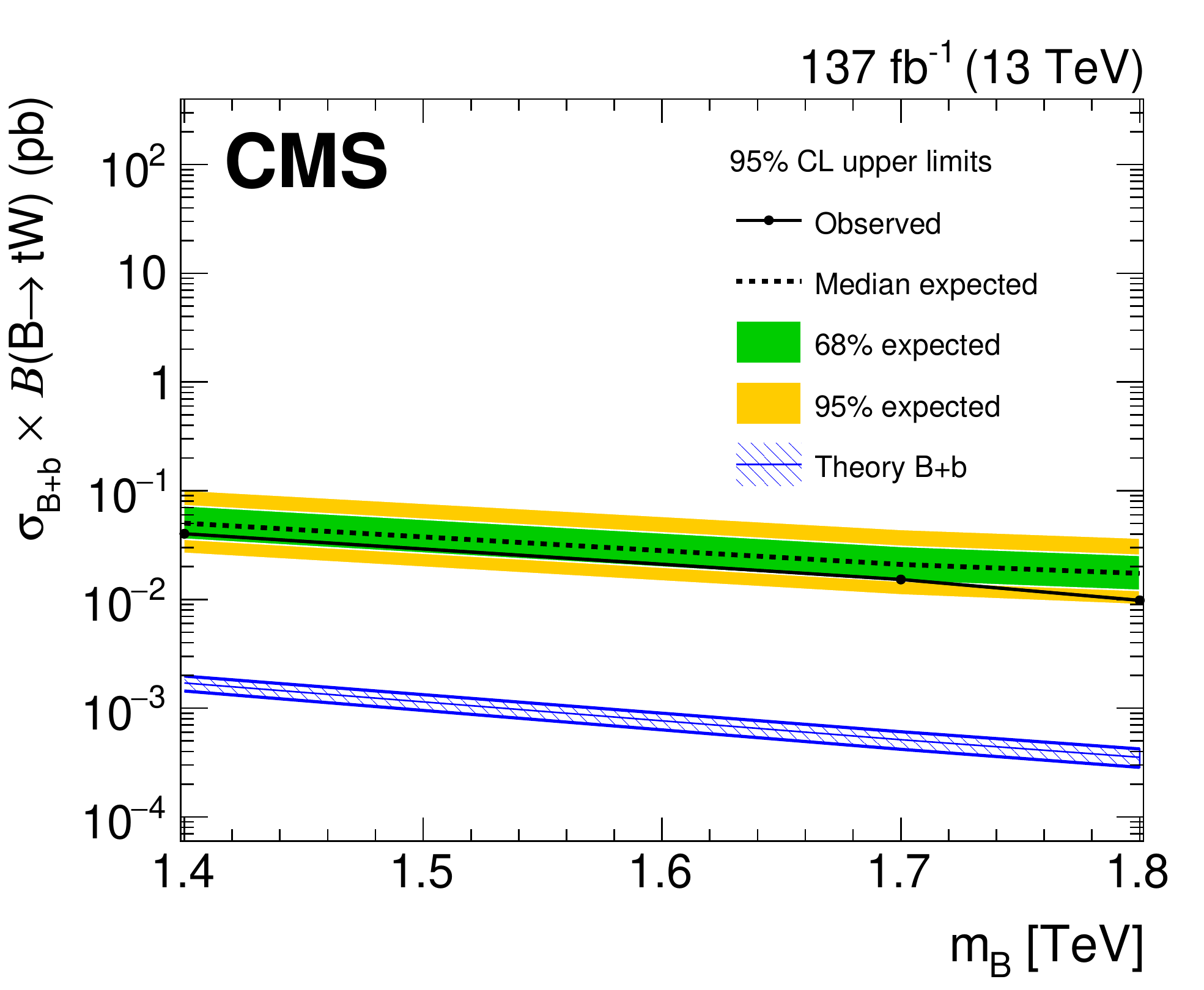}
    \includegraphics[width=0.49\textwidth]{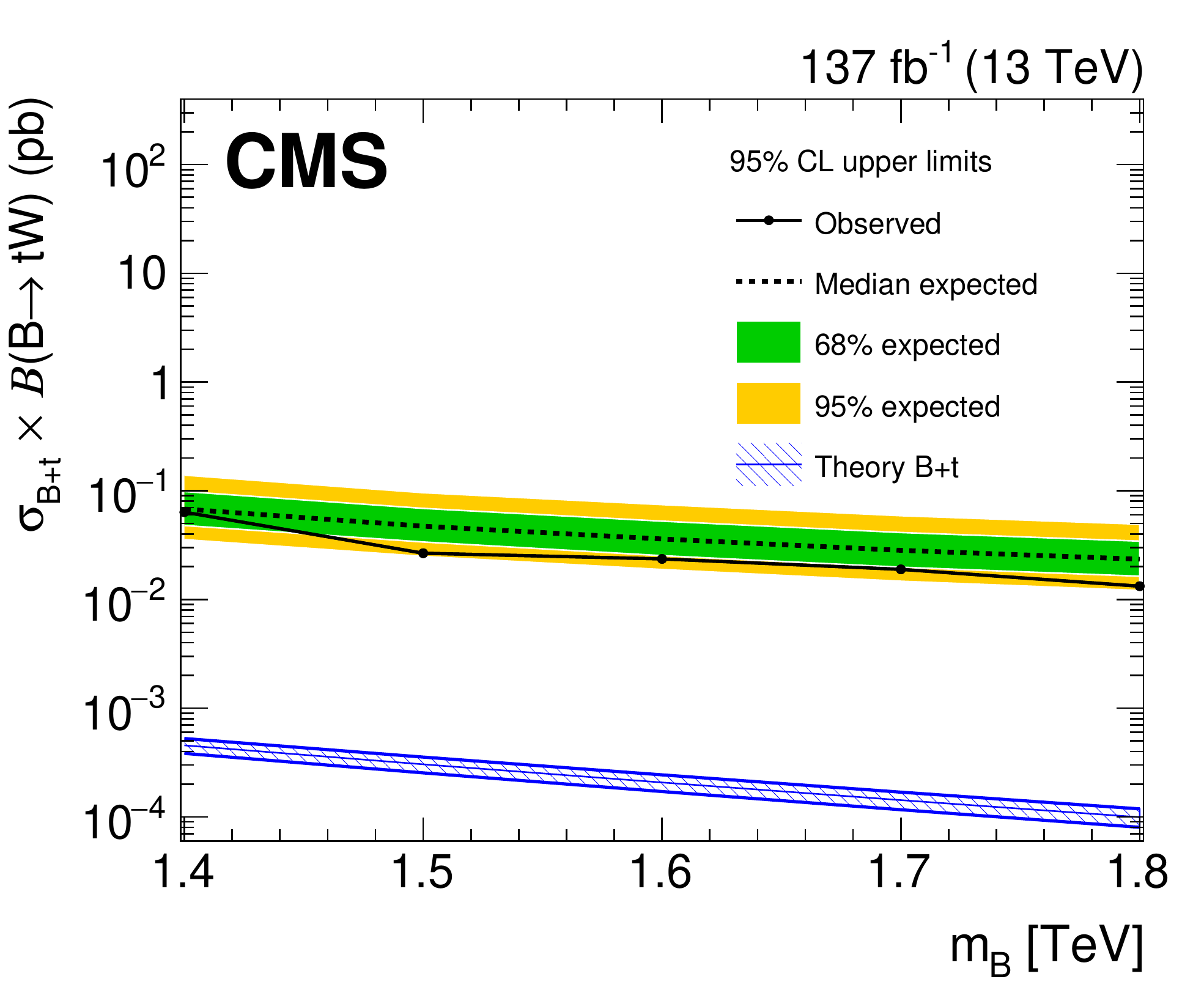}

    \caption{
        Upper limits on the product of the cross section and branching fraction at 95\% $\CL$
        for a $\bpr$ produced in association with a bottom quark (left) and top quark (right)
        as a function of the $\bpr$ quark mass.
        The expected (dashed) and observed (dot-solid) limits, as well as the $\bpr$ quark
        theoretical cross sections (shaded-solid), are shown. 
        The inner and outer shaded areas around the expected limits show the 68\% and 95\% CL intervals, respectively.
        }
    \label{figs:limitsBprime}
\end{figure}

The results of this search can also be used to test models of a single $\bpr$ quark produced via the electroweak interaction
in association with a bottom or top quark, and decaying into a top quark and a \PW boson.
Because the cross section for this process is much smaller than for a \bst quark produced through
the strong force, and because of the selection $\mtw > 1.2 \TeV$, we consider the mass range
1.4 to 1.8\TeV in this interpretation. The exclusion limits on $\sigma_{\bpr} \mathcal{B}(\bpr \to \tw \to \text{hadrons})$ at 95\% \CL
are shown in Fig.~\ref{figs:limitsBprime}.
Over the mass range of 1.4--1.8\TeV, the observed upper limit ranges 0.027 to 0.009\unit{pb}
when produced in association with a bottom quark and from 0.036 to 0.012\unit{pb} when produced in association with a top quark.
Because of the small theoretical cross section for the model considered, no mass limit is set.
When compared to the \bst quark in this mass range, the expected cross section upper limits for a $\bpr$ quark
produced with an associated bottom quark are uniformly more sensitive by approximately 22\%.
The equivalent comparison for a $\bpr$ quark produced with an associated top quark shows the sensitivity is worse by no more than 7\%.

These results based on 137\fbinv of data can be compared directly to those of Ref.~\cite{Bprime2016},
which analyzed the lepton+jets channel in 35.9\fbinv of data recorded with the CMS experiment at $\sqrt{s} =  13\TeV$.
At a $\bpr$ quark mass of 1.4\TeV, this analysis is less sensitive than the results
from Ref.~\cite{Bprime2016} by about 20\% when considering $\bpr$ quark production with an associated top quark.
However, this analysis has about 20\% higher sensitivity than the previous analysis when the production is in association
with a bottom quark. As the $\bpr$ quark mass increases, the sensitivity of this analysis increases faster
than the analysis described in Ref.~\cite{Bprime2016}. Thus, the sensitivity of this analysis at 1.8\TeV
is about 27\% higher for the associated top quark hypothesis and about a factor of two higher for the associated bottom
quark hypothesis. This comparison is also applicable to the results for a left-handed $\bpr$ hypothesis in Ref.~\cite{ATLASbprime},
which analyzed the lepton+jets channel in 36.1\fbinv of data recorded with the ATLAS experiment at $\sqrt{s} =  13\TeV$
and which has comparable results to those in Ref.~\cite{Bprime2016}.

\clearpage
        
\section{Summary}
\label{sec:summary}

A search for a heavy resonance decaying to a top quark and a \PW boson
in the fully hadronic final state has been presented. The analysis uses proton-proton
collision data at a center-of-mass energy of 13\TeV corresponding to an integrated
luminosity of 137\fbinv, collected by the CMS experiment at the LHC during 2016--2018.

This analysis considers the explicit case where the heavy resonance is
an excited bottom quark, \bst.
The search evaluates \bst quark masses greater than 1.2\TeV,
which result in highly Lorentz-boosted top quarks and \PW bosons that
are reconstructed as single jets. Using jet substructure
algorithms designed to distinguish heavy resonance decays
from light-quark and gluon jets, the top quark and \PW boson decays are
identified as a top quark jet and a \PW boson jet, respectively.

The background processes in the analysis are
a result of multijet processes from the strong interaction, \ttbar production, and single top quark (tW-channel) production.
The search is performed using a two-dimensional binned likelihood fit to the
data that allows all backgrounds to be fit simultaneously.
The multijet component in the signal region is estimated via a two-dimensional transfer function
method that uses a multijet-enriched control region.
The \ttbar and single top background estimates are determined via
a template fit to data. In particular, a dedicated \ttbar measurement region
is used to constrain the shape and yield of the \ttbar background. 

No statistically
significant deviation from the standard model expectation is observed.
The hypotheses of \bst quarks with left-handed, right-handed,
and vector-like chiralities are excluded at 95\% confidence level for masses below
2.6, 2.8, and 3.1\TeV, respectively.
These are the most stringent limits
on the \bst quark mass to date, extending the previous
best mass limits by almost a factor of two.

\begin{acknowledgments}
    We congratulate our colleagues in the CERN accelerator departments for the excellent performance of the LHC and thank the technical and administrative staffs at CERN and at other CMS institutes for their contributions to the success of the CMS effort. In addition, we gratefully acknowledge the computing centers and personnel of the Worldwide LHC Computing Grid and other centers for delivering so effectively the computing infrastructure essential to our analyses. Finally, we acknowledge the enduring support for the construction and operation of the LHC, the CMS detector, and the supporting computing infrastructure provided by the following funding agencies: BMBWF and FWF (Austria); FNRS and FWO (Belgium); CNPq, CAPES, FAPERJ, FAPERGS, and FAPESP (Brazil); MES (Bulgaria); CERN; CAS, MoST, and NSFC (China); MINCIENCIAS (Colombia); MSES and CSF (Croatia); RIF (Cyprus); SENESCYT (Ecuador); MoER, ERC PUT and ERDF (Estonia); Academy of Finland, MEC, and HIP (Finland); CEA and CNRS/IN2P3 (France); BMBF, DFG, and HGF (Germany); GSRT (Greece); NKFIA (Hungary); DAE and DST (India); IPM (Iran); SFI (Ireland); INFN (Italy); MSIP and NRF (Republic of Korea); MES (Latvia); LAS (Lithuania); MOE and UM (Malaysia); BUAP, CINVESTAV, CONACYT, LNS, SEP, and UASLP-FAI (Mexico); MOS (Montenegro); MBIE (New Zealand); PAEC (Pakistan); MSHE and NSC (Poland); FCT (Portugal); JINR (Dubna); MON, RosAtom, RAS, RFBR, and NRC KI (Russia); MESTD (Serbia); SEIDI, CPAN, PCTI, and FEDER (Spain); MOSTR (Sri Lanka); Swiss Funding Agencies (Switzerland); MST (Taipei); ThEPCenter, IPST, STAR, and NSTDA (Thailand); TUBITAK and TAEK (Turkey); NASU (Ukraine); STFC (United Kingdom); DOE and NSF (USA).

    \hyphenation{Rachada-pisek} Individuals have received support from the Marie-Curie program and the European Research Council and Horizon 2020 Grant, contract Nos.\ 675440, 724704, 752730, 765710 and 824093 (European Union); the Leventis Foundation; the Alfred P.\ Sloan Foundation; the Alexander von Humboldt Foundation; the Belgian Federal Science Policy Office; the Fonds pour la Formation \`a la Recherche dans l'Industrie et dans l'Agriculture (FRIA-Belgium); the Agentschap voor Innovatie door Wetenschap en Technologie (IWT-Belgium); the F.R.S.-FNRS and FWO (Belgium) under the ``Excellence of Science -- EOS" -- be.h project n.\ 30820817; the Beijing Municipal Science \& Technology Commission, No. Z191100007219010; the Ministry of Education, Youth and Sports (MEYS) of the Czech Republic; the Deutsche Forschungsgemeinschaft (DFG), under Germany's Excellence Strategy -- EXC 2121 ``Quantum Universe" -- 390833306, and under project number 400140256 - GRK2497; the Lend\"ulet (``Momentum") Program and the J\'anos Bolyai Research Scholarship of the Hungarian Academy of Sciences, the New National Excellence Program \'UNKP, the NKFIA research grants 123842, 123959, 124845, 124850, 125105, 128713, 128786, and 129058 (Hungary); the Council of Science and Industrial Research, India; the Ministry of Science and Higher Education and the National Science Center, contracts Opus 2014/15/B/ST2/03998 and 2015/19/B/ST2/02861 (Poland); the National Priorities Research Program by Qatar National Research Fund; the Ministry of Science and Higher Education, project no. 0723-2020-0041 (Russia); the Programa Estatal de Fomento de la Investigaci{\'o}n Cient{\'i}fica y T{\'e}cnica de Excelencia Mar\'{\i}a de Maeztu, grant MDM-2015-0509 and the Programa Severo Ochoa del Principado de Asturias; the Thalis and Aristeia programs cofinanced by EU-ESF and the Greek NSRF; the Rachadapisek Sompot Fund for Postdoctoral Fellowship, Chulalongkorn University and the Chulalongkorn Academic into Its 2nd Century Project Advancement Project (Thailand); the Kavli Foundation; the Nvidia Corporation; the SuperMicro Corporation; the Welch Foundation, contract C-1845; and the Weston Havens Foundation (USA).
\end{acknowledgments}

\bibliography{auto_generated}

\providecommand{\href}[2]{#2}\begingroup\raggedright\begin{thebibliography}{10}%
\makeatletter
\providecommand{\hrefCMSnoop }[0]{\@secondoftwo}%
\makeatother
\providecommand{\doi}{\texttt{doi:}\begingroup \urlstyle{tt}\Url}

\bibitem{PhysRevD.42.815}
\hrefCMSnoop {}{U.~Baur, M.~Spira, and P.~M. Zerwas, ``{Excited-quark and
  -lepton production at hadron colliders}'',} \textit{ Phys. Rev. D} \textbf{
  42} (1990) 815,
  \href{http://dx.doi.org/10.1103/PhysRevD.42.815}{\doi{10.1103/PhysRevD.42.815}}.

\bibitem{Tait:2000sh}
\hrefCMSnoop {}{T.~M.~P. Tait and C.-P. Yuan, ``{Single top quark production as
  a window to physics beyond the standard model}'',} \textit{ Phys. Rev. D}
  \textbf{ 63} (2000) 014018,
  \href{http://dx.doi.org/10.1103/PhysRevD.63.014018}{\doi{10.1103/PhysRevD.63.014018}},
\href{http://www.arXiv.org/abs/hep-ph/0007298}{\texttt{arXiv:hep-ph/0007298}}.
%%CITATION = HEP-PH/0007298;%%.

\bibitem{RS1}
\hrefCMSnoop {}{C.~Cheung, A.~L. Fitzpatrick, and L.~Randall, ``{Sequestering
  CP violation and GIM-violation with warped extra dimensions}'',} \textit{
  JHEP} \textbf{ 01} (2008) 069,
  \href{http://dx.doi.org/10.1088/1126-6708/2008/01/069}{\doi{10.1088/1126-6708/2008/01/069}},
  \href{http://www.arXiv.org/abs/0711.4421}{\texttt{arXiv:0711.4421}}.

\bibitem{RS2}
\hrefCMSnoop {}{A.~L. Fitzpatrick, G.~Perez, and L.~Randall, ``{Flavor anarchy
  in a Randall-Sundrum model with 5D minimal flavor violation and a low
  Kaluza-Klein scale}'',} \textit{ Phys. Rev. Lett.} \textbf{ 100} (2008)
  171604,
  \href{http://dx.doi.org/10.1103/physrevlett.100.171604}{\doi{10.1103/physrevlett.100.171604}},
  \href{http://www.arXiv.org/abs/0710.1869}{\texttt{arXiv:0710.1869}}.

\bibitem{HGP1}
\hrefCMSnoop {}{C.~Bini, R.~Contino, and N.~Vignaroli, ``{Heavy-light decay
  topologies as a new strategy to discover a heavy gluon}'',} \textit{ JHEP}
  \textbf{ 01} (2012) 157,
  \href{http://dx.doi.org/10.1007/jhep01(2012)157}{\doi{10.1007/jhep01(2012)157}},
  \href{http://www.arXiv.org/abs/1110.6058}{\texttt{arXiv:1110.6058}}.

\bibitem{HGP2}
\hrefCMSnoop {}{N.~Vignaroli, ``{Discovering the composite Higgs through the
  decay of a heavy fermion}'',} \textit{ JHEP} \textbf{ 07} (2012) 158,
  \href{http://dx.doi.org/10.1007/jhep07(2012)158}{\doi{10.1007/jhep07(2012)158}},
  \href{http://www.arXiv.org/abs/1204.0468}{\texttt{arXiv:1204.0468}}.

\bibitem{HGP3}
\hrefCMSnoop {}{N.~Vignaroli, ``{$\Delta F = 1$ constraints on composite Higgs
  models with left-right parity}'',} \textit{ Phys. Rev. D} \textbf{ 86} (2012)
  115011,
  \href{http://dx.doi.org/10.1103/physrevd.86.115011}{\doi{10.1103/physrevd.86.115011}},
  \href{http://www.arXiv.org/abs/1204.0478}{\texttt{arXiv:1204.0478}}.

\bibitem{Collaboration_2008}
\hrefCMSnoop {}{{CMS Collaboration}, ``{The CMS experiment at the CERN LHC}'',}
  \textit{ JINST} \textbf{ 3} (2008) S08004.,
  \href{http://dx.doi.org/10.1088/1748-0221/3/08/S08004}{\doi{10.1088/1748-0221/3/08/S08004}}.

\bibitem{Nutter_2012}
\hrefCMSnoop {}{J.~W. Nutter, R.~Schwienhorst, D.~G.~E. Walker, and J.-H. Yu,
  ``{Single top production as a probe of B' quarks}'',} \textit{ Phys. Rev. D}
  \textbf{ 86} (2012) 094006,
  \href{http://dx.doi.org/10.1103/physrevd.86.094006}{\doi{10.1103/physrevd.86.094006}},
  \href{http://www.arXiv.org/abs/1207.5179}{\texttt{arXiv:1207.5179}}.

\bibitem{Aad:2013rna}
\hrefCMSnoop {}{{ATLAS Collaboration}, ``{Search for single
  $\textrm{b}^*$-quark production with the ATLAS detector at $\sqrt{s}=7$
  TeV}'',} \textit{ Phys. Lett. B} \textbf{ 721} (2013) 171,
  \href{http://dx.doi.org/10.1016/j.physletb.2013.03.016}{\doi{10.1016/j.physletb.2013.03.016}},
\href{http://www.arXiv.org/abs/1301.1583}{\texttt{arXiv:1301.1583}}.
%%CITATION = ARXIV:1301.1583;%%.

\bibitem{Khachatryan2016}
\hrefCMSnoop {}{{CMS Collaboration}, ``{Search for the production of an excited
  bottom quark decaying to tW in proton-proton collisions at $\sqrt{s}=8$
  TeV}'',} \textit{ JHEP} \textbf{ 01} (2016) 166,
  \href{http://dx.doi.org/10.1007/JHEP01(2016)166}{\doi{10.1007/JHEP01(2016)166}},
  \href{http://www.arXiv.org/abs/1509.08141}{\texttt{arXiv:1509.08141}}.

\bibitem{bstarTobg}
\hrefCMSnoop {}{{CMS Collaboration}, ``{Search for resonances and quantum black
  holes using dijet mass spectra in proton-proton collisions at $\sqrt{s} =$ 8
  TeV}'',} \textit{ Phys. Rev. D} \textbf{ 91} (2015) 052009,
  \href{http://dx.doi.org/10.1103/PhysRevD.91.052009}{\doi{10.1103/PhysRevD.91.052009}},
  \href{http://www.arXiv.org/abs/1501.04198}{\texttt{arXiv:1501.04198}}.

\bibitem{bstarTobgATLAS}
\hrefCMSnoop {}{{ATLAS Collaboration}, ``{Search for new resonances in mass
  distributions of jet pairs using 139 $fb^{-1}$ of pp collisions at $\sqrt{s}
  =$ 13 TeV with the ATLAS detector}'',} \textit{ JHEP} \textbf{ 03} (2020)
  145,
  \href{http://dx.doi.org/10.1007/jhep03(2020)145}{\doi{10.1007/jhep03(2020)145}},
  \href{http://www.arXiv.org/abs/1910.08447}{\texttt{arXiv:1910.08447}}.

\bibitem{Larkoski_2020}
\hrefCMSnoop {}{A.~J. Larkoski, I.~Moult, and B.~Nachman, ``{Jet substructure
  at the Large Hadron Collider: A review of recent advances in theory and
  machine learning}'',} \textit{ Phys. Rep.} \textbf{ 841} (2020) 1,
  \href{http://dx.doi.org/10.1016/j.physrep.2019.11.001}{\doi{10.1016/j.physrep.2019.11.001}},
  \href{http://www.arXiv.org/abs/1709.04464}{\texttt{arXiv:1709.04464}}.

\bibitem{Kogler_2019}
\hrefCMSnoop {}{R.~Kogler {et~al.}, ``{Jet substructure at the Large Hadron
  Collider}'',} \textit{ Rev. Mod. Phys.} \textbf{ 91} (2019) 045003,
  \href{http://dx.doi.org/10.1103/revmodphys.91.045003}{\doi{10.1103/revmodphys.91.045003}},
  \href{http://www.arXiv.org/abs/1803.06991}{\texttt{arXiv:1803.06991}}.

\bibitem{VLQ1}
\hrefCMSnoop {}{J.~A. Aguilar-Saavedra, R.~Benbrik, S.~Heinemeyer, and
  M.~P{\'ue}rez-Victoria, ``Handbook of vectorlike quarks: Mixing and single
  production'',} \textit{ Phys. Rev. D} \textbf{ 88} (2013) 094010,
  \href{http://dx.doi.org/10.1103/physrevd.88.094010}{\doi{10.1103/physrevd.88.094010}},
  \href{http://www.arXiv.org/abs/1306.0572}{\texttt{arXiv:1306.0572}}.

\bibitem{VLQ2}
\hrefCMSnoop {}{A.~De~Simone, O.~Matsedonskyi, R.~Rattazzi, and A.~Wulzer, ``A
  first top partner hunter's guide'',} \textit{ JHEP} \textbf{ 04} (2013) 004,
  \href{http://dx.doi.org/10.1007/jhep04(2013)004}{\doi{10.1007/jhep04(2013)004}},
  \href{http://www.arXiv.org/abs/1211.5663}{\texttt{arXiv:1211.5663}}.

\bibitem{Khachatryan:2016bia}
\hrefCMSnoop {}{{CMS Collaboration}, ``{The CMS trigger system}'',} \textit{
  JINST} \textbf{ 12} (2017) P01020,
  \href{http://dx.doi.org/10.1088/1748-0221/12/01/P01020}{\doi{10.1088/1748-0221/12/01/P01020}},
\href{http://www.arXiv.org/abs/1609.02366}{\texttt{arXiv:1609.02366}}.
%%CITATION = ARXIV:1609.02366;%%.

\bibitem{Krohn_2010}
\hrefCMSnoop {}{D.~Krohn, J.~Thaler, and L.-T. Wang, ``{Jet Trimming}'',}
  \textit{ JHEP} \textbf{ 02} (2010) 084,
  \href{http://dx.doi.org/10.1007/JHEP02(2010)084}{\doi{10.1007/JHEP02(2010)084}},
  \href{http://www.arXiv.org/abs/0912.1342}{\texttt{arXiv:0912.1342}}.

\bibitem{CMS-PAS-TOP-16-011}
\hrefCMSnoop {}{{CMS Collaboration}, ``{Measurements of
  $\mathrm{t\overline{t}}$ differential cross sections in proton-proton
  collisions at $\sqrt{s}=$ 13 TeV using events containing two leptons}'',}
  \textit{ JHEP} \textbf{ 02} (2019) 149,
  \href{http://dx.doi.org/10.1007/JHEP02(2019)149}{\doi{10.1007/JHEP02(2019)149}},
  \href{http://www.arXiv.org/abs/1811.06625}{\texttt{arXiv:1811.06625}}.

\bibitem{CMS-PAS-TOP-16-008}
\hrefCMSnoop {}{{CMS Collaboration}, ``{Measurement of differential cross
  sections for top quark pair production using the lepton+jets final state in
  proton-proton collisions at 13 TeV}'',} \textit{ Phys. Rev. D} \textbf{ 95}
  (2017) 092001,
  \href{http://dx.doi.org/10.1103/PhysRevD.95.092001}{\doi{10.1103/PhysRevD.95.092001}},
  \href{http://www.arXiv.org/abs/1610.04191}{\texttt{arXiv:1610.04191}}.

\bibitem{Re:2010bp}
\hrefCMSnoop {}{E.~Re, ``{Single-top Wt-channel production matched with parton
  showers using the POWHEG method}'',} \textit{ Eur. Phys. J. C} \textbf{ 71}
  (2011) 1547,
  \href{http://dx.doi.org/10.1140/epjc/s10052-011-1547-z}{\doi{10.1140/epjc/s10052-011-1547-z}},
  \href{http://www.arXiv.org/abs/1009.2450}{\texttt{arXiv:1009.2450}}.

\bibitem{Nason:2004rx}
\hrefCMSnoop {}{P.~Nason, ``{A new method for combining NLO QCD with shower
  Monte Carlo algorithms}'',} \textit{ JHEP} \textbf{ 11} (2004) 040,
  \href{http://dx.doi.org/10.1088/1126-6708/2004/11/040}{\doi{10.1088/1126-6708/2004/11/040}},
  \href{http://www.arXiv.org/abs/hep-ph/0409146}{\texttt{arXiv:hep-ph/0409146}}.

\bibitem{Frixione:2007vw}
\hrefCMSnoop {}{S.~Frixione, P.~Nason, and C.~Oleari, ``{Matching NLO QCD
  computations with parton shower simulations: The POWHEG method}'',} \textit{
  JHEP} \textbf{ 11} (2007) 070,
  \href{http://dx.doi.org/10.1088/1126-6708/2007/11/070}{\doi{10.1088/1126-6708/2007/11/070}},
  \href{http://www.arXiv.org/abs/0709.2092}{\texttt{arXiv:0709.2092}}.

\bibitem{Alioli:2010xd}
\hrefCMSnoop {}{S.~Alioli, P.~Nason, C.~Oleari, and E.~Re, ``{A general
  framework for implementing NLO calculations in shower Monte Carlo programs:
  The POWHEG BOX}'',} \textit{ JHEP} \textbf{ 06} (2010) 043,
  \href{http://dx.doi.org/10.1007/JHEP06(2010)043}{\doi{10.1007/JHEP06(2010)043}},
  \href{http://www.arXiv.org/abs/1002.2581}{\texttt{arXiv:1002.2581}}.

\bibitem{Powhegttbar}
\hrefCMSnoop {}{S.~{Frixione}, G.~{Ridolfi}, and P.~{Nason}, ``{A
  positive-weight next-to-leading-order Monte Carlo for heavy flavour
  hadroproduction}'',} \textit{ JHEP} \textbf{ 09} (2007) 126,
  \href{http://dx.doi.org/10.1088/1126-6708/2007/09/126}{\doi{10.1088/1126-6708/2007/09/126}},
  \href{http://www.arXiv.org/abs/0707.3088}{\texttt{arXiv:0707.3088}}.

\bibitem{Alwall:2014hca}
J.~Alwall\hrefCMSnoop {}{ {et~al.}, ``{The automated computation of tree-level
  and next-to-leading order differential cross sections, and their matching to
  parton shower simulations}'',} \textit{ JHEP} \textbf{ 07} (2014) 079,
  \href{http://dx.doi.org/10.1007/JHEP07(2014)079}{\doi{10.1007/JHEP07(2014)079}},
  \href{http://www.arXiv.org/abs/1405.0301}{\texttt{arXiv:1405.0301}}.

\bibitem{Sjostrand:2014zea}
T.~Sj{\"o}strand\hrefCMSnoop {}{ {et~al.}, ``{An introduction to PYTHIA
  8.2}'',} \textit{ Comput. Phys. Commun.} \textbf{ 191} (2015) 159,
  \href{http://dx.doi.org/10.1016/j.cpc.2015.01.024}{\doi{10.1016/j.cpc.2015.01.024}},
  \href{http://www.arXiv.org/abs/1410.3012}{\texttt{arXiv:1410.3012}}.

\bibitem{Ball_2015}
\hrefCMSnoop {}{{NNPDF} Collaboration, ``{Parton distributions for the LHC Run
  II}'',} \textit{ JHEP} \textbf{ 04} (2015) 040,
  \href{http://dx.doi.org/10.1007/JHEP04(2015)040}{\doi{10.1007/JHEP04(2015)040}},
  \href{http://www.arXiv.org/abs/1410.8849}{\texttt{arXiv:1410.8849}}.

\bibitem{CUETP8M1}
\hrefCMSnoop {}{{CMS Collaboration}, ``{Event generator tunes obtained from
  underlying event and multiparton scattering measurements}'',} \textit{ Eur.
  Phys. J. C} \textbf{ 76} (2016) 155,
  \href{http://dx.doi.org/10.1140/epjc/s10052-016-3988-x}{\doi{10.1140/epjc/s10052-016-3988-x}},
  \href{http://www.arXiv.org/abs/1512.00815}{\texttt{arXiv:1512.00815}}.

\bibitem{Ball_2017}
\hrefCMSnoop {}{{NNPDF} Collaboration, ``{Parton distributions from
  high-precision collider data}'',} \textit{ Eur. Phys. J. C} \textbf{ 77}
  (2017) 663,
  \href{http://dx.doi.org/10.1140/epjc/s10052-017-5199-5}{\doi{10.1140/epjc/s10052-017-5199-5}},
  \href{http://www.arXiv.org/abs/1706.00428}{\texttt{arXiv:1706.00428}}.

\bibitem{CP5}
\hrefCMSnoop {}{{CMS Collaboration}, ``{Extraction and validation of a new set
  of CMS PYTHIA8 tunes from underlying-event measurements}'',} \textit{ Eur.
  Phys. J. C} \textbf{ 80} (2020) 4,
  \href{http://dx.doi.org/10.1140/epjc/s10052-019-7499-4}{\doi{10.1140/epjc/s10052-019-7499-4}},
  \href{http://www.arXiv.org/abs/1903.12179}{\texttt{arXiv:1903.12179}}.

\bibitem{GEANT4}
\hrefCMSnoop {}{{GEANT4} Collaboration, ``{\GEANTfour}---a simulation
  toolkit'',} \textit{ Nucl. Instrum. Meth. A} \textbf{ 506} (2003) 250,
\href{http://dx.doi.org/10.1016/S0168-9002(03)01368-8}{\doi{10.1016/S0168-9002(03)01368-8}}.
%%CITATION = NUIMA,A506,250;%%.

\bibitem{Aaboud:2016mmw}
\hrefCMSnoop {}{{ATLAS Collaboration}, ``{Measurement of the inelastic
  proton-proton cross section at $\sqrt{s} = 13$ TeV with the ATLAS Detector at
  the LHC}'',} \textit{ Phys. Rev. Lett.} \textbf{ 117} (2016) 182002,
  \href{http://dx.doi.org/10.1103/PhysRevLett.117.182002}{\doi{10.1103/PhysRevLett.117.182002}},
  \href{http://www.arXiv.org/abs/1606.02625}{\texttt{arXiv:1606.02625}}.

\bibitem{Sirunyan:2018nqx}
\hrefCMSnoop {}{{CMS Collaboration}, ``{Measurement of the inelastic
  proton-proton cross section at $ \sqrt{s}=13 $ TeV}'',} \textit{ JHEP}
  \textbf{ 07} (2018) 161,
  \href{http://dx.doi.org/10.1007/JHEP07(2018)161}{\doi{10.1007/JHEP07(2018)161}},
  \href{http://www.arXiv.org/abs/1802.02613}{\texttt{arXiv:1802.02613}}.

\bibitem{Cacciari:2008gp}
\hrefCMSnoop {}{M.~Cacciari, G.~P. Salam, and G.~Soyez, ``{The anti-\kt jet
  clustering algorithm}'',} \textit{ JHEP} \textbf{ 04} (2008) 063,
  \href{http://dx.doi.org/10.1088/1126-6708/2008/04/063}{\doi{10.1088/1126-6708/2008/04/063}},
  \href{http://www.arXiv.org/abs/0802.1189}{\texttt{arXiv:0802.1189}}.

\bibitem{Cacciari:2011ma}
\hrefCMSnoop {}{M.~Cacciari, G.~P. Salam, and G.~Soyez, ``{FastJet user
  manual}'',} \textit{ Eur. Phys. J. C} \textbf{ 72} (2012) 1896,
  \href{http://dx.doi.org/10.1140/epjc/s10052-012-1896-2}{\doi{10.1140/epjc/s10052-012-1896-2}},
\href{http://www.arXiv.org/abs/1111.6097}{\texttt{arXiv:1111.6097}}.
%%CITATION = ARXIV:1111.6097;%%.

\bibitem{CMS-PRF-14-001}
\hrefCMSnoop {}{{CMS Collaboration}, ``{Particle-flow reconstruction and global
  event description with the CMS detector}'',} \textit{ JINST} \textbf{ 12}
  (2017) P10003,
  \href{http://dx.doi.org/10.1088/1748-0221/12/10/P10003}{\doi{10.1088/1748-0221/12/10/P10003}},
\href{http://www.arXiv.org/abs/1706.04965}{\texttt{arXiv:1706.04965}}.
%%CITATION = ARXIV:1706.04965;%%.

\bibitem{Bertolini:2014bba}
\hrefCMSnoop {}{D.~Bertolini, P.~Harris, M.~Low, and N.~Tran, ``{Pileup per
  particle identification}'',} \textit{ JHEP} \textbf{ 10} (2014) 059,
  \href{http://dx.doi.org/10.1007/JHEP10(2014)059}{\doi{10.1007/JHEP10(2014)059}},
\href{http://www.arXiv.org/abs/1407.6013}{\texttt{arXiv:1407.6013}}.
%%CITATION = ARXIV:1407.6013;%%.

\bibitem{Sirunyan:2020foa}
\hrefCMSnoop {}{{CMS Collaboration}, ``{Pileup mitigation at CMS in 13 TeV
  data}'',} \textit{ JINST} \textbf{ 15} (2020) P09018,
  \href{http://dx.doi.org/10.1088/1748-0221/15/09/p09018}{\doi{10.1088/1748-0221/15/09/p09018}},
  \href{http://www.arXiv.org/abs/2003.00503}{\texttt{arXiv:2003.00503}}.

\bibitem{Khachatryan:2016kdb}
\hrefCMSnoop {}{{CMS Collaboration}, ``{Jet energy scale and resolution in the
  CMS experiment in pp collisions at 8 TeV}'',} \textit{ JINST} \textbf{ 12}
  (2017) P02014,
  \href{http://dx.doi.org/10.1088/1748-0221/12/02/P02014}{\doi{10.1088/1748-0221/12/02/P02014}},
\href{http://www.arXiv.org/abs/1607.03663}{\texttt{arXiv:1607.03663}}.
%%CITATION = ARXIV:1607.03663;%%.

\bibitem{CMS-PAS-JME-16-003}
\href {http://cds.cern.ch/record/2256875}{{CMS Collaboration}, ``Jet algorithms
  performance in 13 {TeV} data'',} CMS Physics Analysis Summary
  CMS-PAS-JME-16-003, 2017.

\bibitem{Larkoski:2014wba}
\hrefCMSnoop {}{A.~J. Larkoski, S.~Marzani, G.~Soyez, and J.~Thaler, ``Soft
  drop'',} \textit{ JHEP} \textbf{ 05} (2014) 146,
  \href{http://dx.doi.org/10.1007/JHEP05(2014)146}{\doi{10.1007/JHEP05(2014)146}},
\href{http://www.arXiv.org/abs/1402.2657}{\texttt{arXiv:1402.2657}}.
%%CITATION = ARXIV:1402.2657;%%.

\bibitem{Dasgupta:2013ihk}
\hrefCMSnoop {}{M.~Dasgupta, A.~Fregoso, S.~Marzani, and G.~P. Salam, ``Towards
  an understanding of jet substructure'',} \textit{ JHEP} \textbf{ 09} (2013)
  029,
  \href{http://dx.doi.org/10.1007/JHEP09(2013)029}{\doi{10.1007/JHEP09(2013)029}},
\href{http://www.arXiv.org/abs/1307.0007}{\texttt{arXiv:1307.0007}}.
%%CITATION = ARXIV:1307.0007;%%.

\bibitem{Butterworth:2008iy}
\hrefCMSnoop {}{J.~M. Butterworth, A.~R. Davison, M.~Rubin, and G.~P. Salam,
  ``{Jet substructure as a new Higgs search channel at the LHC}'',} \textit{
  Phys. Rev. Lett.} \textbf{ 100} (2008) 242001,
  \href{http://dx.doi.org/10.1103/PhysRevLett.100.242001}{\doi{10.1103/PhysRevLett.100.242001}},
\href{http://www.arXiv.org/abs/0802.2470}{\texttt{arXiv:0802.2470}}.
%%CITATION = ARXIV:0802.2470;%%.

\bibitem{Thaler:2010tr}
\hrefCMSnoop {}{J.~Thaler and K.~Van~Tilburg, ``{Identifying boosted objects
  with {$N$}-subjettiness}'',} \textit{ JHEP} \textbf{ 03} (2011) 015,
  \href{http://dx.doi.org/10.1007/JHEP03(2011)015}{\doi{10.1007/JHEP03(2011)015}},
\href{http://www.arXiv.org/abs/1011.2268}{\texttt{arXiv:1011.2268}}.
%%CITATION = ARXIV:1011.2268;%%.

\bibitem{Thaler_2012}
\hrefCMSnoop {}{J.~Thaler and K.~Van~Tilburg, ``{Maximizing boosted top
  identification by minimizing $N$-subjettiness}'',} \textit{ JHEP} \textbf{
  02} (2012) 093,
  \href{http://dx.doi.org/10.1007/JHEP02(2012)093}{\doi{10.1007/JHEP02(2012)093}},
  \href{http://www.arXiv.org/abs/1108.2701}{\texttt{arXiv:1108.2701}}.

\bibitem{Sirunyan:2017ezt}
\hrefCMSnoop {}{{CMS Collaboration}, ``{Identification of heavy-flavour jets
  with the CMS detector in pp collisions at 13 TeV}'',} \textit{ JINST}
  \textbf{ 13} (2018) P05011,
  \href{http://dx.doi.org/10.1088/1748-0221/13/05/P05011}{\doi{10.1088/1748-0221/13/05/P05011}},
  \href{http://www.arXiv.org/abs/1712.07158}{\texttt{arXiv:1712.07158}}.

\bibitem{CMS-PAS-JME-18-002}
\hrefCMSnoop {}{{CMS Collaboration}, ``{Identification of heavy, energetic,
  hadronically decaying particles using machine-learning techniques}'',}
  \textit{ JINST} \textbf{ 15} (2020) P06005,
  \href{http://dx.doi.org/10.1088/1748-0221/15/06/P06005}{\doi{10.1088/1748-0221/15/06/P06005}},
  \href{http://www.arXiv.org/abs/2004.08262}{\texttt{arXiv:2004.08262}}.

\bibitem{PDG}
\hrefCMSnoop {}{{Particle Data Group}, P.~A. Zyla {et~al.}, ``Review of
  particle physics'',} \textit{ Prog. Theor. Exp. Phys.} \textbf{ 2020} (2020)
  083C01,
  \href{http://dx.doi.org/10.1093/ptep/ptaa104}{\doi{10.1093/ptep/ptaa104}}.

\bibitem{10.2307/2340521}
\hrefCMSnoop {}{R.~A. Fisher, ``{On the interpretation of $\chi^2$ from
  contingency tables, and the calculation of p}'',} \textit{ J. Royal Stat.
  Soc.} \textbf{ 85} (1922) 87,
  \href{http://dx.doi.org/10.2307/2340521}{\doi{10.2307/2340521}}.

\bibitem{Cranmer2000KernelEI}
\hrefCMSnoop {}{K.~S. Cranmer, ``{Kernel estimation in high-energy physics}'',}
  \textit{ Comput. Phys. Commun.} \textbf{ 136} (2001) 198,
  \href{http://dx.doi.org/10.1016/S0010-4655(00)00243-5}{\doi{10.1016/S0010-4655(00)00243-5}},
  \href{http://www.arXiv.org/abs/hep-ex/0011057}{\texttt{arXiv:hep-ex/0011057}}.

\bibitem{CMS-PAS-LUM-17-001}
\href {https://cds.cern.ch/record/2257069}{{CMS Collaboration}, ``{CMS
  luminosity measurements for the 2016 data taking period}'',} CMS Physics
  Analysis Summary CMS-PAS-LUM-17-001, 2017.

\bibitem{CMS-PAS-LUM-17-004}
\href {https://cds.cern.ch/record/2621960}{{CMS Collaboration}, ``{CMS
  luminosity measurement for the 2017 data-taking period at $\sqrt{s} =
  13~\mathrm{TeV}$}'',} CMS Physics Analysis Summary CMS-PAS-LUM-17-004, 2018.

\bibitem{CMS-PAS-LUM-18-002}
\href {https://cds.cern.ch/record/2676164}{{CMS Collaboration}, ``{CMS
  luminosity measurement for the 2018 data-taking period at $\sqrt{s} =
  13~\mathrm{TeV}$}'',} CMS Physics Analysis Summary CMS-PAS-LUM-18-002, 2019.

\bibitem{Butterworth_2016}
\hrefCMSnoop {}{J.~Butterworth {et~al.}, ``{PDF4LHC recommendations for LHC Run
  II}'',} \textit{ J. Phys. G} \textbf{ 43} (2016) 023001,
  \href{http://dx.doi.org/10.1088/0954-3899/43/2/023001}{\doi{10.1088/0954-3899/43/2/023001}},
  \href{http://www.arXiv.org/abs/1510.03865}{\texttt{arXiv:1510.03865}}.

\bibitem{Baker:1983tu}
\hrefCMSnoop {}{S.~Baker and R.~D. Cousins, ``{Clarification of the use of chi
  square and likelihood functions in fits to histograms}'',} \textit{ Nucl.
  Instrum. Meth.} \textbf{ 221} (1984) 437,
  \href{http://dx.doi.org/10.1016/0167-5087(84)90016-4}{\doi{10.1016/0167-5087(84)90016-4}}.

\bibitem{parametricInference}
J.~K. Lindsey, ``Parametric statistical inference''.
\newblock Oxford University Press, New York, 1966.

\bibitem{Cowan_2011}
\hrefCMSnoop {}{G.~Cowan, K.~Cranmer, E.~Gross, and O.~Vitells, ``Asymptotic
  formulae for likelihood-based tests of new physics'',} \textit{ Eur. Phys. J.
  C} \textbf{ 71} (2011) 1554,
  \href{http://dx.doi.org/10.1140/epjc/s10052-011-1554-0}{\doi{10.1140/epjc/s10052-011-1554-0}},
  \href{http://www.arXiv.org/abs/1007.1727}{\texttt{arXiv:1007.1727}}.
  [Erratum: \DOI{10.1140/epjc/s10052-013-2501-z}].

\bibitem{CMSDijet}
\hrefCMSnoop {}{{CMS Collaboration}, ``{Search for high mass dijet resonances
  with a new background prediction method in proton-proton collisions at
  $\sqrt{s} = 13~\mathrm{TeV}$}'',} \textit{ JHEP} \textbf{ 05} (2020) 033,
  \href{http://dx.doi.org/10.1007/jhep05(2020)033}{\doi{10.1007/jhep05(2020)033}},
  \href{http://www.arXiv.org/abs/1911.03947}{\texttt{arXiv:1911.03947}}.

\bibitem{Bprime2016}
\hrefCMSnoop {}{{CMS Collaboration}, ``{Search for single production of
  vector-like quarks decaying to a top quark and a W boson in proton-proton
  collisions at $\sqrt{s} =$ 13 TeV}'',} \textit{ Eur. Phys. J. C} \textbf{ 79}
  (2019) 90,
  \href{http://dx.doi.org/10.1140/epjc/s10052-019-6556-3}{\doi{10.1140/epjc/s10052-019-6556-3}},
  \href{http://www.arXiv.org/abs/1809.08597}{\texttt{arXiv:1809.08597}}.

\bibitem{ATLASbprime}
\hrefCMSnoop {}{{ATLAS Collaboration}, ``{Search for pair production of heavy
  vector-like quarks decaying into high-$\pt$ W bosons and top quarks in the
  lepton-plus-jets final state in pp collisions at $\sqrt{s}=$ 13 TeV with the
  ATLAS detector}'',} \textit{ JHEP} \textbf{ 08} (2018) 048,
  \href{http://dx.doi.org/10.1007/jhep08(2018)048}{\doi{10.1007/jhep08(2018)048}},
  \href{http://www.arXiv.org/abs/1806.01762}{\texttt{arXiv:1806.01762}}.

\end{thebibliography}\endgroup
\cleardoublepage \appendix\section{The CMS Collaboration \label{app:collab}}\begin{sloppypar}\hyphenpenalty=5000\widowpenalty=500\clubpenalty=5000\vskip\cmsinstskip
\textbf{Yerevan Physics Institute, Yerevan, Armenia}\\*[0pt]
A.M.~Sirunyan$^{\textrm{\dag}}$, A.~Tumasyan
\vskip\cmsinstskip
\textbf{Institut f\"{u}r Hochenergiephysik, Wien, Austria}\\*[0pt]
W.~Adam, T.~Bergauer, M.~Dragicevic, A.~Escalante~Del~Valle, R.~Fr\"{u}hwirth\cmsAuthorMark{1}, M.~Jeitler\cmsAuthorMark{1}, N.~Krammer, L.~Lechner, D.~Liko, I.~Mikulec, F.M.~Pitters, J.~Schieck\cmsAuthorMark{1}, R.~Sch\"{o}fbeck, M.~Spanring, S.~Templ, W.~Waltenberger, C.-E.~Wulz\cmsAuthorMark{1}, M.~Zarucki
% ...
\vskip\cmsinstskip
\dag: Deceased\\
1:  Also at TU Wien, Wien, Austria\\
% ...
\end{sloppypar}
%%% END EDITABLE REGION %%%
% skeleton_end
\end{document}